\documentclass[10pt]{iopart}
\usepackage{iopams}
\usepackage{color}
\usepackage{graphicx}
\usepackage{natbib}
\usepackage{bm}
\usepackage{ulem}

\def\vts{v_{{\rm t}s}}

\def\vti{v_{{\rm ti}}}

\def\rhoLsO{\rho_{{\rm c}0s}}

\def\omLs{\omega_{{\rm c}s}}
\def\omLsO{\omega_{{\rm c}0s}}

\def\omLi{\omega_{{\rm ci}}}
\def\omLiO{\omega_{{\rm c0i}}}

\def\omegad{\omega_{\rm d}}
\def\omegads{\omega_{{\rm d}s}}
\def\omegadi{\omega_{\rm di}}

\def\vAO{v_{{\rm A0}}}

\def\omAO{\omega_{{\rm A0}}}

\def\me{m_{\rm e}}
\def\mi{m_{\rm i}}

\def\Ts{\mathcal{T}_s}
\def\Te{\mathcal{T}_{\rm e}}
\def\Ti{\mathcal{T}_{\rm i}}

\def\tauei{\tau_{\rm ei}}

\def\feqs{f_{0s}}
\def\feqi{f_{0{\rm i}}}

\def\FMs{F_{{\rm M}s}}
\def\FMi{F_{\rm Mi}}
\def\nsO{n_{s0}}

\def\niO{n_{{\rm i}0}}

\def\L{\mathcal{L}}
\def\T{\mathcal{T}}
\def\O{\mathcal{O}}
\def\enr{\mathcal{E}}

\def\dphi{\delta\phi}
\def\dpsi{\delta\psi}
\def\dBp{\delta B_\parallel}

\def\eps{\varepsilon}

\def\sgnVp{\hat{\sigma}}

\def\krhoi{\hat{k}_{\rm i}}

\def\krhoiO{\hat{k}_{\rm 0i}}
\def\krhosO{\hat{k}_{{\rm 0}s}}

\def\nablab{\bm\nabla}

\definecolor{gray}{rgb}{0.5,0.5,0.5}
\definecolor{dred}{rgb}{0.5,0.0,0.0}
\definecolor{dgreen}{rgb}{0.0,0.5,0.0}
\definecolor{dblue}{rgb}{0.0,0.0,0.5}

\def\newblock{\hskip .11em plus .33em minus .07em}

\begin{document}

\title[$\alpha$TAE: II. Kinetic excitation with ITG]{Pressure-gradient-induced Alfv\'{e}n eigenmodes: \\ II. Kinetic excitation with ion temperature gradient}

\author{A Bierwage$^{1,\footnotemark}$, L Chen$^{1,2}$ and F Zonca$^3$}
\address{$^1$ Department of Physics and Astronomy, University of California, \\ Irvine, CA 92697, USA}
\address{$^2$ Institute for Fusion Theory and Simulation, Zhejiang University, \\ Hangzhou, People's Republic of China}
\address{$^3$ Associazione EURATOM-ENEA sulla Fusione, CP 65-00044 Frascati, \\ Rome, Italy}
\eads{\mailto{andreas@bierwage.de} \mailto{liuchen@uci.edu} \mailto{zonca@frascati.enea.it}}

\footnotetext{Present address: Associazione EURATOM-ENEA sulla Fusione, CP 65-00044 Frascati, Rome, Italy.}


\begin{abstract}
The kinetic excitation of ideal magnetohydrodynamic (MHD) discrete Alfv\'{e}n eigenmodes in the second MHD ballooning stable domain is studied in the presence of a thermal ion temperature gradient (ITG), using linear gyrokinetic particle-in-cell simulations of a local flux tube in shifted-circle tokamak geometry. The instabilities are identified as $\alpha$-induced toroidal Alfv\'{e}n eigenmodes ($\alpha$TAE); that is, bound states trapped between pressure-gradient-induced potential barriers of the Schr\"{o}dinger equation for shear Alfv\'{e}n waves. Using numerical tools, we examine in detail the effect of kinetic thermal ion compression on $\alpha$TAEs; both non-resonant coupling to ion sound waves and wave-particle resonances. It is shown that the Alfv\'{e}nic ITG instability thresholds (e.g., the critical temperature gradient) are determined by two resonant absorption mechanisms: Landau damping and continuum damping. The numerical results are interpreted on the basis of a theoretical framework previously derived from a variational formulation. The present analysis of properties and structures of Alfv\'{e}nic fluctuations in the presence of steep pressure gradients applies for both positive or negative magnetic shear and can serve as an interpretative framework for experimental observations in (future) high-performance fusion plasmas of reactor relevance.
\end{abstract}


\section{Introduction}
\label{sec:intro}

Ideal magnetohydrodynamic (MHD) theory for a tokamak plasma predicts local stability with respect to MHD ballooning modes around flux surfaces with positive magnetic shear when the pressure gradient exceeds a certain threshold and in the entire domain of negative magnetic shear \cite{Coppi80}. The prospect of utilizing this property to achieve high plasma pressures in devices designed to create thermonuclear fusion conditions has raised the question how this so-called second MHD ballooning stable domain is affected by kinetic effects, such as wave-particle interactions. Studies focusing on the effect of thermal ion temperature gradient (ITG) have shown differing results; some workers found a destabilization of virtually the entire ideal MHD stable domain \cite{Hirose94, Hirose95, Nordman95, Yamagiwa97, Joiner08}, others only a kinetic broadening of the ideal MHD ballooning unstable domain \cite{Dong04}. A destabilization was also observed in the presence of an electron temperature gradient (ETG) \cite{Dong04, Joiner07}. These numerical gyrokinetic studies focused primarily on the question of stability or instability, but comparatively little insight was obtained with regard to the origin of the eigenfrequencies, which is important to understand the nature of the underlying eigenmodes. In another recent study, it was demonstrated that, in the second MHD stable domain, discrete ideal MHD Alfv\'{e}n eigenmodes exist \cite{Hu04}, which may be resonantly excited, for instance, through interactions with a sparse energetic ion population \cite{Hu05}. One may now ask the question whether the ITG instabilities reported in \cite{Hirose94, Hirose95, Nordman95, Yamagiwa97, Joiner08, Dong04} are related to the discrete Alfv\'{e}n eigenmodes described in \cite{Hu04, Hu05}.

These ideal MHD eigenmodes, which are known as $\alpha$-induced toroidal Alfv\'{e}n eigenmodes ($\alpha$TAE), are solutions of the Schr\"{o}dinger equation for shear Alfv\'{e}n waves (SAW), which are trapped between potential barriers induced by the pressure gradient. There is an infinite number of potential barriers and, thus, an infinite number of branches, which are denoted as $\alpha$TAE$(j,p)$. The label $j\geq 1$ identifies the potential well where the dominant component is trapped, and $p\geq 0$ is the energy quantum number. In those parameter regimes where the eigenfrequency of a given $\alpha$TAE branch is comparable to the diamagnetic frequency $\omega_{*p{\rm i}}$ and about three times the thermal ion transit frequency, $3\times\omega_{\rm ti}$,\footnote{The factor of $3$ is due to the fact that the distribution function is weighted towards high energies by a factor $\enr^{5/2}$, where $\enr$ is the kinetic energy of a particle (details are given in section \protect\ref{sec:wpi_particles}).} coupling to ion sound waves and resonant interaction with thermal ions is expected to have a significant effect.

The purpose of the present paper is to examine in detail such interactions between $\alpha$TAEs and thermal ions in the presence of an ITG. The topics dealt with in the present paper may be summarized as follows:
\begin{enumerate}
\item  New interpretation of ITG-driven instabilities reported in \cite{Hirose94, Dong04} in terms of $\alpha$TAEs \cite{Hu04, Hu05}. (based on results for frequencies and mode structures obtained in the companion paper \cite{Bierwage10a})
\item  What determines which branch of $\alpha$TAEs dominates? (highlight role of minimized field line bending)
\item  Which factors determine the mode frequency in the kinetic simulations? (analysis with respect to field line bending, diamagnetic drift and kinetic compression)
\item  Which parts of the particle distribution drive the instability, which ones damp it? (examine response of individual particles in pressure-curvature coupling term)
\item  How can we relate our results to a theoretical framework based on a variational analysis? (relation between numerical results without scale separation and theoretical results based on a two-scale local mode structure)
\end{enumerate}

We utilize a reduced version of the linear gyrokinetic equations derived by Chen \& Hasegawa \cite{Chen91}, which describe the dynamics of drift Alfv\'{e}n ballooning modes and treat electrons as a massless fluid. The gyrokinetic equations are solved with the linear $\delta f$ particle-in-cell (PIC) code \textsc{awecs} \cite{Bierwage08}. Following the approach chosen in related earlier studies of the second MHD stable domain (e.g., \cite{Hirose94, Dong04, Hu04}), we employ the shifted-circle model equilibrium, which describes tokamak-type configurations in terms of two parameters: the average magnetic shear $s$ and the normalized pressure gradient $\alpha$ \cite{Connor78}. An isotropic velocity distribution is used and the modulation of the magnetic field strength $B$ is ignored, so $B = B_0$, there are no magnetically trapped particles and no toroidicity-induced Alfv\'{e}n frequency gap. However, magnetic curvature and $\nablab B$ drifts are properly accounted for. Parameters are adopted from \cite{Hirose94, Dong04}, so that we may revisit the cases considered there. The model and methods used are described in section \ref{sec:model}.

In this setting, a kinetic destabilization of the second MHD stable domain is seen as in \cite{Hirose94, Hirose95, Nordman95, Yamagiwa97, Joiner08}. Based on the results given in a companion paper \cite{Bierwage10a}, where the effect of finite Larmor radii (FLR) on $\alpha$TAEs is examined in the fluid limit, we are able to identify all kinetic instabilities observed in the second MHD stable domain as $\alpha$TAEs destabilized via interaction with thermal ions. This is shown in section \ref{sec:atae_comp}.

In section~\ref{sec:qforms} and \ref{sec:wpi}, we analyze the simulation results from the viewpoint of non-resonant and resonant dynamics of kinetic thermal ion compression. The discussion in section~\ref{sec:discuss} is used to understand numerical results and explain the observations on the basis of a theoretical framework derived from a variational formulation, which was incrementally developed in \cite{Chen84, Tsai93, Chen94, Zonca96, Zonca99, Zonca06b, Chen07, Zonca07a} (see \cite{Chen07} for a review). In section~\ref{sec:conclude}, the results are summarized and conclusions are drawn.

\section{Physical model and numerical methods}
\label{sec:model}

We employ the linear $\delta f$ gyrokinetic particle-in-cell code \textsc{awecs} introduced in \cite{Bierwage08}, which solves a set of one-dimensional linear gyrokinetic equations based on the derivation by Chen \& Hasegawa \cite{Chen91} for a local flux tube described by the ballooning formalism in terms of the extended poloidal coordinate $-\infty < \theta < \infty$ \cite{Connor78, Coppi77, Lee77, Pegoraro78, Dewar81}. The high-$\beta$ tokamak-like flux tube geometry is approximated using the well-known shifted-circle model equilibrium, which is parametrized in terms of the average magnetic shear $s$ and normalized pressure gradient $\alpha$ \cite{Connor78}. The electromagnetic field perturbations are described by gyrokinetic Maxwell equations in terms of the magnetic flux function $\delta\psi$, the electrostatic potential $\delta\phi$, and the magnetic compression $\delta B_\parallel$. The linear gyrokinetic Vlasov equation governs the evolution of the fluctuating part, $\delta f_s$, of the total distribution function, $f_s = \feqs + \delta f_s$. Thermal ions ($s={\rm i}$) and electrons ($s={\rm e}$) are assumed to be described by an (isotropic) Maxwellian distribution, $\feqs = \nsO\FMs$. Electrons are approximated by a massless fluid, and the contributions from energetic ions are neglected in the present work. The resulting equations describe linear drift Alfv\'{e}n ballooning modes subject to kinetic thermal ion compression. A detailed description of the model and the numerical methods is given in \cite{Bierwage08}. Our formulation makes use of the following coefficients and parameters:
\begin{eqnarray}
\fl & Q_s = \frac{\omega_{*s}^T - \omega}{\T_s}, \quad 
\omega_{*s}^T = \omega_{*s}\left[1 + \eta_s\left(\frac{\enr}{\T_s} - \frac{3}{2}\right)\right], \quad
\omega_{*ps} = \omega_{*s}(1 + \eta_s), &
\label{eq:defs}
\\
\fl & \omega_{*s} = \frac{k_\vartheta\T_s}{\omLsO L_n}, \quad
\eta_s = \frac{\T_s'/\T_s}{\nsO'/\nsO}, \quad
L_n^{-1} = -\frac{\nsO'}{\nsO}, \quad
\eps_n = \frac{L_n}{R_0}, \quad
\tau_{{\rm e}s}^T = \frac{\me\Te}{m_s\T_s}, & \nonumber
\\
\fl & \omegads = \frac{\Omega_\kappa}{\omLsO} \left(v_\parallel^2 + \mu B_0\right) + \frac{\Omega_p}{\omLsO} \mu B_0, \quad
\Omega_\kappa = \frac{k_\vartheta}{R_0} g, \quad
\Omega_p = -\frac{k_\vartheta \alpha}{2 q^2 R_0}, \quad
\omLs = e_s B/m_s, & \nonumber
\\
\fl & k_\perp = \sqrt{f} k_\vartheta, \quad
\krhosO = k_\vartheta \vts/\omLsO, \quad
\lambda_s = k_\perp \rhoLsO = \sqrt{f} \krhoiO v_\perp / \vts, \quad
b_{0s} = f \krhosO^2, & \nonumber
\\
\fl & f = 1 + h^2, \quad
g = \cos\theta + h\sin\theta, \quad
h = s(\theta-\theta_k) - \alpha\sin\theta. & \nonumber
\end{eqnarray}

\noindent Much of the notation is standard and definitions of all quantities can be found in a companion paper \cite{Bierwage10a} where the (FLR MHD) fluid limit of the model is analyzed. In the present study, we focus on modes with constant radial envelope ($\theta_k = 0$).

\subsection{Gyrokinetic equations}
\label{sec:gk}

Following standard procedures, the adiabatic and convective responses are separated through the substitution
\begin{equation}
\delta f_s = -\frac{e_s\feqs}{m_s\Ts}\left(\dphi + \frac{Q_s\T_s}{\omega} J_0\dpsi e^{iL_k}\right) + \frac{\delta G_s}{\omega} e^{iL_k};
\label{eq:model_vlasov_gke1}
\end{equation}

\noindent where $L_k = -{\bm k}_\perp\cdot(\hat{\bm b}\times{\bm v}_\perp)/\omLsO$. The quantity $\delta G_s$ captures the compressional part of the non-adiabatic component of the particle response (in short, kinetic compression). Electrons are approximated as a massless fluid, so that $\delta G_{\rm e} = 0$. The evolution of kinetic ion compression, $\delta G_{\rm i}$, is governed by the gyrokinetic equation \cite{Chen91},
\begin{equation}
\left[\frac{v_\parallel}{q R_0} \partial_\theta - i(\omega - \omegadi) \right] \delta G_{\rm i} = i \frac{e_{\rm i}\feqi}{\mi} Q_{\rm i} (\delta S_{1{\rm i}} - i\sgnVp \delta S_{2{\rm i}});
\label{eq:model_vlasov_gke2}
\end{equation}

\noindent where $\sgnVp = v_\parallel/|v_\parallel|$ and the source terms are
\begin{eqnarray}
& \delta S_{1{\rm i}} = J_0 (\delta\phi - \delta\psi) + \omegadi J_0 \dpsi + \frac{v_\perp}{k_\perp} J_1 \omega \dBp, &
\\
& \delta S_{2{\rm i}} = -\frac{|v_\parallel|}{q R_0}(\partial_\theta\lambda_{\rm i}) J_1 \dpsi. &
\label{eq:gke_ds2}
\end{eqnarray}

\noindent The gyrokinetic field equations for isotropic $\feqi$ and written in terms of $\delta G_{\rm i}$ and in Laplace-transformed form ($\partial_t \rightarrow -i\omega$, with $\omega = \omega_{\rm r} + i\gamma$) read \cite{Bierwage08}
\begin{eqnarray}
\fl 0 =& \underbrace{\frac{k_\vartheta^2}{(q R_0)^2} \frac{\partial}{\partial\theta}\left(f \frac{\partial\delta\psi}{\partial\theta}\right)}\limits_{\rm FLB}
\underbrace{- \mu_0 \left<\frac{e^2}{m}(1 - J_0^2) Q f_0 \right>_{\rm i} \omega\delta\phi}\limits_{\rm inertia\; (ideal\; MHD\; +\; FLR)}
\underbrace{- \Omega_p\left[(\Omega_p + 2\Omega_\kappa)\delta\psi + \omega\delta B_\parallel\right]}\limits_{\rm MPC\; +\; MFC\; (drift{\tt -}kinetic)}
\label{eq:model_maxw_vort1}
\\
\fl & \underbrace{- \mu_0\left<\frac{e^2}{m}\omegad(1 - J_0^2) Q f_0 \right>_{\rm i} \delta\psi
- \frac{\mu_0}{B}\left<e \mu B \left( 1 - \frac{2J_1}{\lambda} J_0\right) Q f_0 \right>_{\rm i} \omega\delta B_\parallel}\limits_{\rm MPC\; +\; MFC\; (FLR\; correction)} \nonumber
\\
\fl & \underbrace{- \mu_0 \left<e\omegad J_0\delta G \right>_{\rm i}
+ i\mu_0 \left<e\frac{v_\parallel}{q R_0}(\partial_\theta\lambda) J_1\delta G \right>_{\rm i},}\limits_{\rm KPC} \nonumber
\\
\fl 0 =& \left<e J_0\delta G \right>_{\rm i}
+ \left<\frac{e^2}{m} \partial_{\mathcal E} f_0\right>_{\rm i} \omega(\delta\phi - \delta\psi) 
+ \left<\frac{e^2}{m} (1 - J_0^2) Q f_0\right>_{\rm i} \delta\psi,
\label{eq:model_maxw_qn1}
\\
\fl 0 =& \omega\delta B_\parallel + \Omega_p\delta\psi
+ \frac{\mu_0}{B} \left<e\mu B\left(1 - \frac{2J_1}{\lambda}J_0\right) Q f_0\right>_{\rm i} \delta\psi
+ \frac{\mu_0}{B} \left<m\mu B \frac{2J_1}{\lambda}\delta G \right>_{\rm i};
\label{eq:model_maxw_bp1}
\end{eqnarray}

\noindent where $\left<...\right> = \int{\rm d}^3 v = \sum_{\sgnVp}\int{\rm d}\enr\int{\rm d}\mu B/|v_\parallel|$ is the velocity space integral. The vorticity equation (\ref{eq:model_maxw_vort1}) consists of the following terms: field line bending (FLB), inertia (with FLR), MHD particle compression (MPC), magnetic field compression (MFC), and kinetic particle compression (KPC). Both MPC and MFC are static compression effects associated with toroidal curvature and finite $\beta$, whereas KPC captures dynamic compression. Equation (\ref{eq:model_maxw_qn1}) is the quasi-neutrality condition, equation (\ref{eq:model_maxw_bp1}) is the perpendicular Amp\`{e}re's law, and both contain additional kinetic compression terms.

For given parameter values, only the fastest growing mode is captured due to the initial value simulation approach. Frequencies and growth rates are normalized by the Alfv\'{e}n frequency $\omega_{\rm A0} = \vAO/(qR_0)$, and velocities by the Alfv\'{e}n velocity $\vAO = B_0/\sqrt{\mu_0 m_{\rm i}n_{\rm 0i}}$.

\subsection{Reduced model for numerical analysis of shear Alfv\'{e}n eigenmodes}
\label{sec:saw}

Being designed for a wide range of parameters, \textsc{awecs} solves, by default, the full equations (\ref{eq:model_vlasov_gke2}) and (\ref{eq:model_maxw_vort1})--(\ref{eq:model_maxw_bp1}). Before carrying out detailed analyses of the results obtained, it is useful to reduce the complexity of the model by eliminating negligible contributions.

Our ordering analyses and numerical studies show that the following approximations can be made in the gyrokinetic Maxwell-Vlasov equations (\ref{eq:model_vlasov_gke2}) and (\ref{eq:model_maxw_vort1})--(\ref{eq:model_maxw_bp1}) at the expense of only quantitative changes when shear Alfv\'{e}n eigenmodes are of interest:
\begin{itemize}
\item  Let $\omega\delta B_\parallel = -\Omega_p\delta\psi$, which effectively eliminates $\Omega_p$ (high-$\beta$ correction to the curvature drift) from $\omegadi$ at the order considered, so that the magnetic drift is given by $\omegadi = \Omega_\kappa(v_\parallel^2 + \mu B_0)/\omLiO$.

\item  Terms containing the factor $v_\parallel(\partial_\theta\lambda_{\rm i}) J_1$ (ion perturbation to the toroidal current) can be ignored.

\item  Although the parallel electric field, $\delta E_\parallel = -\partial_\theta(\dphi - \dpsi)$, influences the frequency and growth rate quantitatively, its presence does not affect Alfv\'{e}n eigenmodes; so, for the purpose of simplicity, its contribution may be ignored.
\end{itemize}

\noindent For convenience, the following dimensionless quantities are used:
\begin{eqnarray}
\hat{\omega} = \frac{\omega}{\omAO}, \quad
\hat{\omega}_{*{\rm i}} = \frac{\omega_{*{\rm i}}}{\omAO} = \frac{q \krhoiO \hat{v}_{\rm ti}}{\eps_n}, \quad
\hat{\Omega}_\kappa = \frac{q R_0 \Omega_\kappa \Ti}{\omLi\vAO} = q \krhoiO \hat{v}_{\rm ti} g(\theta),
\label{eq:norm_omgf_omgk}
\\
\hat{v} = \frac{v}{\vAO}, \quad
\hat{\omega}_{\rm ti}^2 = \hat{v}_{\rm ti}^2 = \frac{\vti^2}{\vAO^2} = \frac{\beta_{\rm i}}{2} = \frac{\alpha \eps_n}{2 q^2 [1 + \eta_{\rm i} + \tauei^T(1 + \eta_{\rm e})]}; \nonumber
\end{eqnarray}

\noindent where $\omega_{\rm A0} = \vAO/(qR_0)$ is the Alfv\'{e}n frequency. In the following, the hats will be neglected on all quantities except $\krhoiO$. For completeness, we retained the contributions due to $\delta E_\parallel$, which may be readily dropped at a later stage by letting $\tauei^T \rightarrow 0$. The resulting \textit{reduced model for drift Alfv\'{e}n ballooning modes subject to kinetic compression} may be compactly written in the form of a Schr\"{o}dinger equation,
\begin{equation}
\delta\Psi_{\rm s}'' - V_{\rm eff}(\theta) \delta\Psi_{\rm s} = 0.
\label{eq:model_saw_red}
\end{equation}

\noindent Here, $\delta\Psi_{\rm s} = \sqrt{f} \delta\psi$, $f = (k_\perp/k_\vartheta)^2 = 1 + h^2$, and $\delta\Psi_{\rm s}'' \equiv {\rm d}^2(\delta\Psi_{\rm s})/{\rm d}\theta^2$. The effective Schr\"{o}dinger potential is written as
\begin{equation}
V_{\rm eff} = V + V_{{\rm m},\omega} + V_{{\rm m},\tau} + V_{\kappa,{\rm FLR}} + V_{\rm ki}.
\label{eq:model_saw_v_eff}
\end{equation}

\noindent The ideal MHD potential, $V$, contributions from the inertia term, $V_{{\rm m},\omega}$ and $V_{{\rm m},\tau}$, and the FLR correction for the ``MPC+MFC'' term, $V_{\kappa,{\rm FLR}}$, were introduced and discussed in \cite{Bierwage10a}. The contribution from the KPC term in (\ref{eq:model_maxw_vort1}) is captured by $V_{\rm ki}$, which is the only addition made here to the FLR MHD equation (7) of \cite{Bierwage10a}:
\begin{equation}
V_{\rm ki}\delta\Psi_{\rm s} = -\frac{\mu_0\vAO^2}{k_\vartheta^2\sqrt{f}} \frac{e\left<\omegad J_0 \delta G\right>_{\rm i}}{\omAO}
= -\frac{\sqrt{f}}{b_{\rm 0i}}\left<\omegad J_0\right. \delta \hat{G}\left>_{\rm i}\right..
\label{eq:model_saw_vflr_kpc}
\end{equation}

\noindent The evolution of $\delta\hat{G}_{\rm i} = \delta G_{\rm i} \times \mi\Ti/(e\niO\omAO)$ is governed by the reduced gyrokinetic equation
\begin{equation}
\left[v_\parallel \partial_\theta - i(\omega - \omegadi) \right] \delta \hat{G}_{\rm i} = i (\omega_{*{\rm i}}^T -\omega)\FMi J_0 (\omegadi\delta\Psi_{\rm s} + \delta U_{\rm s}) / \sqrt{f},
\label{eq:model_vlasov_red}
\end{equation}

\noindent and the contribution of the parallel electric field, represented by $\delta U_{\rm s} = \omega \sqrt{f} (\delta\phi - \delta\psi)$, is given by the quasi-neutrality condition in the form
\begin{equation}
\fl (1 + \tauei^T) \delta U_{\rm s} = \tauei^T\left[\omega_{*{\rm i}}(1 - \Gamma_0\Upsilon_1) - \omega(1 - \Gamma_0)\right] \delta\Psi_{\rm s} - \tauei^T\sqrt{f}\left<J_0\right.\delta \hat{G}\left>_{\rm i}.\right.
\label{eq:model_qn}
\end{equation}

\noindent The linear time-independent Schr\"{o}dinger equation (\ref{eq:model_saw_red}) is used in section \ref{sec:qforms} to define quadratic forms, and in section \ref{sec:discuss} to make a connection to a theoretical framework based on a variational formulation.

\subsection{Parameter values}
\label{sec:model_parms}

In the present paper, we analyze two cases: one with lower magnetic shear ($s=0.4$) and one with higher magnetic shear ($s=1.0$). The default parameters are listed in table \ref{tab:parms}. They were adopted (with minor changes) from two earlier linear gyrokinetic studies on Alfv\'{e}nic ITG-driven instabilities \cite{Hirose94, Dong04}. In the case adopted from Hirose \textit{et al.} \protect\cite{Hirose94}, $\krhoiO$ is increased from $0.1$ to $0.2$ in order to demonstrate the excitation of the $\alpha$TAE(1,0) branch besides the (2,0) branch. The normalized wavenumber used by Dong \textit{et al.} \protect\cite{Dong04} is adjusted by a factor $\sqrt{2}$ due to a different definition of $\vti$ ($\krhoiO = 0.212 = 0.3/\sqrt{2}$).

Note that is is not meaningful to simply compare the values of parameters $(s,\alpha)$ of the shifted-circle equilibrium model used here with parameter values in experimental configurations with non-circular flux surfaces. The location of the stability boundaries of the ideally unstable domain and the distribution of the $\alpha$TAE bands in the $s$-$\alpha$ plane are sensitive to the flux surface geometry, so it is likely that a given region in parameter space of the simple model may correspond to another region of the parameter space in a more realistic geometry. A meaningful comparison would need to consider the shape of the Schr\"{o}dinger potential at different points in parameter space, rather than the values of parameters like $s$ and $\alpha$. To our knowledge, this has not been done yet, so we have no basis for making comparisons with experiments at this point.

\begin{table}
\caption{Default physical parameters in the two cases considered in this paper: one with lower magnetic shear ($s=0.4$) and one with higher shear ($s=1.0$).}
\label{tab:parms}
\begin{indented}
\item[]
\begin{tabular}{@{}ccccccccl}
\br
$s$ & $q$ & $\eps_n$ & $\krhoiO$ & $\eta_{\rm i}$ & $\eta_{\rm e}$ & $\tauei^T$ & $\beta_{\rm i} = \beta_{\rm e}$ & adapted from \\
\mr
$0.4$ & $1.2$ & $0.175$ & $0.2$   & $2$   & $2$   & $1$ & $\approx \alpha/50$ & Hirose~\textit{et al.}~\protect\cite{Hirose94} \\
$1.0$ & $1.5$ & $0.2$   & $0.212$ & $2.5$ & $2.5$ & $1$ & $\approx \alpha/80$ & Dong~\textit{et al.}~\protect\cite{Dong04} \\
\br
\end{tabular}
\end{indented}
\end{table}

\section{Gyrokinetic simulation results: ITG-driven $\alpha$TAEs}
\label{sec:atae_comp}

In this section, results from linear gyrokinetic simulations using \textsc{awecs} \cite{Bierwage08} are presented. We adopt two cases considered previously by Hirose \textit{et al.}~\cite{Hirose94} and Dong \textit{et al.}~\cite{Dong04}. The physical parameters are shown in table \ref{tab:parms}. The two cases are used to highlight similarities and differences between results obtained for lower and higher values of the magnetic shear ($s=0.4$ and $s=1.0$, respectively). In section \ref{sec:scans_alpha}, we present results for parameter scans with respect to the normalized pressure gradient, $\alpha$, which provides an image of the various branches of shear Alfv\'{e}n eigenmodes destabilized in the $s$-$\alpha$ plane. In section \ref{sec:scans_scaling}, parameter scans with respect to the temperature gradient and perpendicular wavenumber, $\eta_{\rm i}$ and $\krhoiO$, are inspected.

In the following, the frequencies $\omega_{\rm r}$ and growth rates $\gamma$ are computed by averaging over a suitable time interval. The associated standard deviations, $\Delta\omega_{\rm r}$ and $\Delta\gamma$, are shown as error bars, which indicate the level of phase-space discretization noise. The results shown are obtained for initial perturbations with mixed parity (i.e., asymmetric around $\theta=0$).

The numerical parameters are chosen as follows. The number of phase-space markers is $N_{\rm m} = 2048\times 13$ (i.e., 26624 markers distributed over 13 periods, $\theta \in [-13\pi,13\pi]$).\footnote{In fact, except near ideal MHD marginal stability, the results in the $s=1.0$ case can be accurately reproduced with markers loaded in as little as 3 periods, $\theta \in [-3\pi,3\pi]$, since $\alpha$TAE eigenvalues are determined by the bound state components only. For the $s=0.4$ case, 5 periods may be sufficient.} There are $N_{\rm g} = 2048$ grid points in the simulation domain $\theta \in [-\theta_{\rm max},\theta_{\rm max}]$ with $\theta_{\rm max} = 120$, and the time step for the 4th-order Runge-Kutta solver is $\Delta t = 0.005 \omAO^{-1}$. Convergence tests were carried out to ensure acceptable numerical accuracy \cite{Bierwage08}. The role of boundary conditions was discussed in the Appendix of \cite{Bierwage10a}.

\begin{figure}[tbp]
\includegraphics[width=1.0\textwidth]
{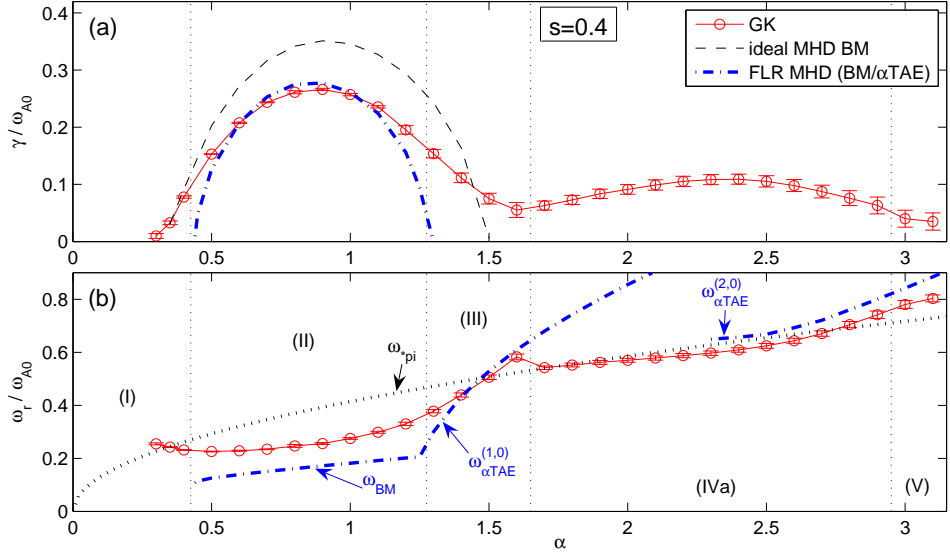}
\caption{Low shear case, $s=0.4$. $\alpha$-dependence of (a) the growth rate $\gamma$ and (b) the frequency $\omega_{\rm r}$. For comparison with the gyrokinetic (GK) simulation results of \textsc{awecs} (\opencircle), (a) shows the growth rate of ballooning modes (BM) in ideal MHD (\broken) and FLR MHD (\chain), and (b) the frequencies $\omega_{\rm BM}$, $\omega_{\alpha{\rm TAE}}^{(1,0)}$ and $\omega_{\alpha{\rm TAE}}^{(2,0)}$ of BMs and $\alpha$TAEs obtained with the FLR MHD model (\chain) \protect\cite{Bierwage10a}. In addition, the diamagnetic frequency $\omega_{*p{\rm i}} \propto \sqrt{\alpha}$ (\dotted) is plotted in (b). The thermal transit frequency, $\omega_{\rm ti}$, is about a factor 5 smaller than $\omega_{*p{\rm i}}$ [cf.~equation (\ref{eq:norm_omgf_omgk})]. The dotted vertical lines separate five regions, labelled (I)--(V), where different branches of Alfv\'{e}n eigenmodes dominate.}
\label{fig:scan-a_hirose}%
\end{figure}

\begin{figure}[tbp]
\includegraphics[width=1.0\textwidth]
{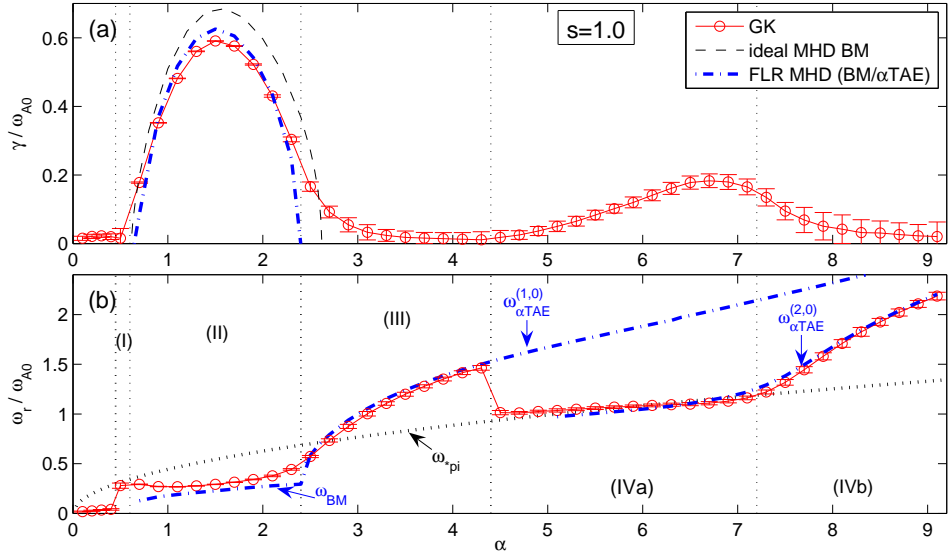}
\caption{High shear case, $s=1.0$. $\alpha$-dependence of (a) the growth rate $\gamma$ and (b) the frequency $\omega_{\rm r}$. Arranged as figure \protect\ref{fig:scan-a_hirose}. Here, the thermal transit frequency, $\omega_{\rm ti}$, is about a factor 4 smaller than $\omega_{*p{\rm i}}$.}
\label{fig:scan-a_dong}%
\end{figure}

\subsection{Pressure gradient scan}
\label{sec:scans_alpha}

The results obtained by scanning the parameter $\alpha$ through the first into the second MHD stable domain are shown in figures~\ref{fig:scan-a_hirose} and \ref{fig:scan-a_dong} for the two cases in table \ref{tab:parms}, $s=0.4$ and $s=1.0$, respectively. In the domain $\alpha < 0.4$ of figure \ref{fig:scan-a_dong}, electrostatic ITG-driven modes ($|\delta\phi| \gg |\delta\psi|$) are dominant. These were described in \cite{Dong92} and will not be considered here. The domain where Alfv\'{e}nic instabilities ($|\delta\psi| \gg |\delta\phi - \delta\psi|$) are observed is divided into five regions, labelled (I)--(V) in figures~\ref{fig:scan-a_hirose} and \ref{fig:scan-a_dong}. In the following paragraphs, the results for regions (I)--(V) are described and discussed one-by-one. Results obtained with the FLR MHD model in \cite{Bierwage10a} are used to identify the various branches of eigenmodes excited by wave-particle interactions.

\paragraph{Region (I):} Alfv\'{e}nic instabilities are found near the first MHD ballooning stability boundary, $0.3 \lesssim \alpha \lesssim 0.4$ in figure \ref{fig:scan-a_hirose}, and $0.5 \lesssim \alpha \lesssim 0.6$ in figure \ref{fig:scan-a_dong}. The frequencies are comparable to the diamagnetic frequency, $\omega_{\rm r} \sim \omega_{*p{\rm i}}$, and mode structures (not shown) are very extended in $\theta$ \cite{Zonca96}. Despite the large simulation domain used in this study ($\theta_{\rm max}=120$), the results are not numerically converged with respect to $\theta_{\rm max}$. The convergence test presented in figure~11 in \cite{Bierwage08} indicates that both $\omega_{\rm r}$ and $\gamma$ may not be accurate (presumably lower). Thus, we are not yet able to determine the critical $\alpha$ value for the onset of Alv\'{e}nic instability below the first MHD ballooning stability boundary and defer a detailed study of these modes. Note that results from earlier studies also differ \cite{Dong04, Dong99, Zhao02}. This regime should be studied with a model including electron inertia and/or finite collisionality in order to resolve the short radial scales which Alfv\'{e}n eigenmodes near ideal MHD marginal stability exhibit (and for which the ion polarization current vanishes).

\paragraph{Region (II):} In the FLR MHD limit \cite{Bierwage10a}, unstable ballooning modes (BM) are found in region (II): $0.45 \lesssim \alpha \lesssim 1.3$ in figure \ref{fig:scan-a_hirose}, and $0.6 \lesssim \alpha \lesssim 2.4$ in figure \ref{fig:scan-a_dong}. Due to the stabilizing effect of FLR terms \cite{Tang81}, this domain is smaller than the ideal MHD ballooning unstable domain, which extends over the range $0.4 \lesssim \alpha \lesssim 1.5$ in figure \ref{fig:scan-a_hirose}, and $0.6 \lesssim \alpha \lesssim 2.6$ in figure \ref{fig:scan-a_dong}.

Away from FLR MHD marginal stability, the growth rates in region (II) are similar for both the gyrokinetic simulation ($\gamma$, \opencircle) and FLR MHD ($\gamma$, \chain). The frequencies in the gyrokinetic simulation ($\omega_{\rm r}$, \opencircle) are larger than in FLR MHD. This up-shift is a manifestation of non-resonant kinetic compression effects. The results in figures \ref{fig:scan-a_hirose} and figure \ref{fig:scan-a_dong} are also consistent with the expectation that the effect of kinetic thermal ion compression (both resonant and non-resonant) becomes increasingly important towards the FLR MHD ballooning stability boundaries where field line bending and ballooning-interchange drive tend to balance \cite{Zonca96}.

\paragraph{Region (III):} Near the second FLR MHD stability boundary, ballooning instabilities modified by kinetic compression smoothly connect to ITG-driven $\alpha$TAE$(1,0)$. This branch provides the dominant instability in region (III), which extends over the range $1.3 \lesssim \alpha \lesssim 1.6$ in figure \ref{fig:scan-a_hirose}, and $2.4 \lesssim \alpha \lesssim 4.6$ in figure \ref{fig:scan-a_dong}. We claim that this is an $\alpha$TAE(1,0), because of the frequency obtained with the gyrokinetic simulation ($\omega_{\rm r}$, \opencircle) matches that found with the FLR MHD model ($\omega_{\alpha{\rm TAE}}^{(1,0)}$, \chain) \cite{Bierwage10a}. This similarity indicates that coupling to ion sound waves has only a small effect on the frequency of the $\alpha$TAE(1,0) branch in region (III), which is confirmed in section \ref{sec:qforms} below. The mode structure of the instability (not shown) peaks in the central potential well ($j=1$) and closely resembles that of the corresponding FLR MHD solution for an $\alpha$TAE(1,0) (figure 3 in \cite{Bierwage10a}). In region (III), the $\alpha$TAE(1,0) branch is quasi-marginally stable (i.e., continuum damping is negligible) \cite{Bierwage10a}, so we conjecture that the decrease in the growth rate $\gamma(\alpha)$ with increasing $\alpha$ is due to a reduction in the resonant drive as field line bending and, thus, $\omega_{\rm r}(\alpha)$, increases faster than $\omega_{*p{\rm i}}(\alpha) \propto \vti(\alpha)$. This detunes the resonance.

\begin{figure}[tbp]
\includegraphics[width=1.0\textwidth]
{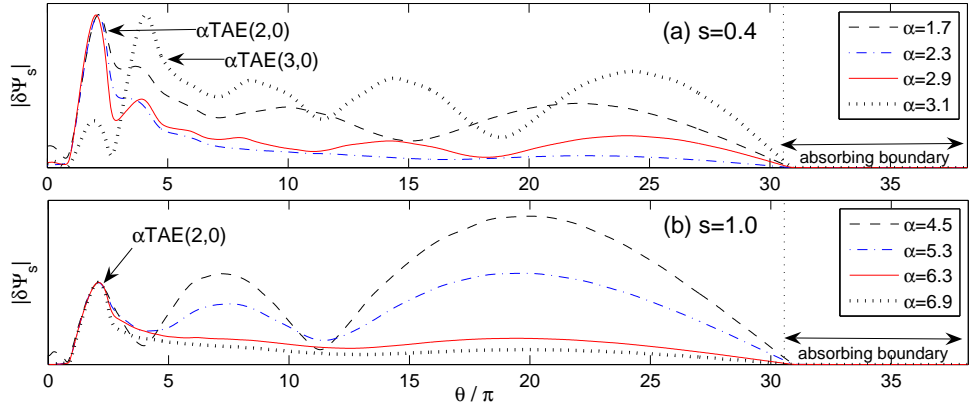}
\caption{Mode structures $|\delta\Psi_{\rm s}(\theta)|$ of ITG-driven $\alpha$TAEs for several values of $\alpha$. (a): $s=0.4$, modes in regions (IVa) and (V) of figure \protect\ref{fig:scan-a_hirose}. (b): $s=1.0$, modes in region (IVa) of figure \protect\ref{fig:scan-a_dong}.}
\label{fig:mstruc_a}%
\end{figure}

\paragraph{Region (IVa):} After a sharp drop in the frequency, the ITG-driven $\alpha$TAE(2,0) branch becomes the dominant instability in region (IVa), which extends over the range $1.7 \lesssim \alpha \lesssim 2.9$ in figure \ref{fig:scan-a_hirose}, and $4.4 \lesssim \alpha \lesssim 7.2$ in figure \ref{fig:scan-a_dong}. We claim that this is an $\alpha$TAE(2,0), because the frequency obtained with the gyrokinetic simulation ($\omega_{\rm r}$, \opencircle) matches that found with the FLR MHD model ($\omega_{\alpha{\rm TAE}}^{(2,0)}$, \chain) \cite{Bierwage10a}.

As was explained in \cite{Bierwage10a}, our analysis of the FLR MHD model allows to identify $\alpha$TAE branches only above a certain numerical threshold, $\alpha_{\rm num}$, which is why the corresponding curves, $\omega_{\alpha{\rm TAE}}^{(2,0)}(\alpha)$ in figures \ref{fig:scan-a_hirose} and \ref{fig:scan-a_dong}, do not reach the low-$\alpha$ boundary of region (IVa). The fact that $\omega_{\rm r}$ in region (IVa) scales like $\alpha^{1/2}$ shows that, in this regime, the $\alpha$TAE(2,0) is below the threshold $\alpha_0^{(2,0)}$ for quasi-marginal stability as defined in \cite{Bierwage10a}. This is the type of modes first seen by Hirose \textit{et al.} \cite{Hirose94}.

The mode structure of the instability, which is shown in figure~\ref{fig:mstruc_a} and peaks in the second potential well ($j=2$), also closely resembles that of the corresponding FLR MHD solution for an $\alpha$TAE(2,0) (figure 4 in \cite{Bierwage10a}). Although the mode structures in figure~\ref{fig:mstruc_a} are plotted only for $\theta > 0$, the modes in region (IV) have mixed parity around $\theta=0$. This shows that kinetic effects preserve the degeneracy (for $\theta_k = 0$) of odd and even solutions of ideal MHD $\alpha$TAE$(j,p)$ branches with $j>1$.

\paragraph{Region (IVb) in figure \ref{fig:scan-a_dong}.} As $\alpha$ increases past the boundary between regions (IVa) and (IVb) in figure \ref{fig:scan-a_dong}, $\alpha \approx 7.2$, the frequency obtained with the gyrokinetic simulation ($\omega_{\rm r}$, \opencircle) continues to follow the $\alpha$TAE(2,0) frequency obtained with the FLR MHD model ($\omega_{\alpha{\rm TAE}}^{(2,0)}$, \chain). The mode structure also continues to resemble an $\alpha$TAE$(2,0)$, as can be seen in figure \ref{fig:mstruc_a}(b). Although the growth rate $\gamma$ decreases with increasing $\alpha$, the amplitude of the large-$\theta$ tail of the mode structure also decreases compared to that of the bound state component ($\pi \lesssim |\theta| \lesssim 3\pi$). This is consistent with the results in figure 6 of \cite{Bierwage10a} and clearly shows that the $\alpha$TAE(2,0) branch becomes quasi-marginally stable in region (IVb); i.e., continuum damping drops to a negligible level when $\alpha > \alpha_0^{(2,0)} \approx 7.2$.

\paragraph{Region (V) in figure \ref{fig:scan-a_hirose}.} As $\alpha$ increases past the boundary between regions (IV) and (V) in figure \ref{fig:scan-a_hirose}, the bound state component $(j,p)=(3,0)$ becomes dominant and the $(2,0)$ component subdominant, as can be seen in figure \ref{fig:mstruc_a}(a). Accordingly, the frequency obtained in the gyrokinetic simulation ($\omega_{\rm r}$, \opencircle) tends to diverge from the shooting code result for the FLR MHD limit ($\omega_{\alpha{\rm TAE}}^{(2,0)}$, \chain) which is locked onto the $\alpha$TAE$(2,0)$ branch. The transition to the $\alpha$TAE$(3,0)$ branch occurs below the point where the $(2,0)$ branch becomes quasi-marginally stable, so there is no region (IVb) here. In the parameter range scanned, it was not possible to find the $\alpha$TAE$(3,0)$ branch with the FLR MHD shooting code, presumably because the corresponding numerical threshold is too large, $\alpha_{\rm num} > 3.2$.

\paragraph{Discussion regarding continuum damping.} As in the example shown in figure \ref{fig:mstruc_a}, the mode structures in regions (III), (IVa) and (V) all exhibit long tails at large $|\theta|$ which correspond to outward propagating FLR continuum waves \cite{Bierwage10a}, a part of which gets reflected at the unphysical boundaries of the simulation domain (cf.~Appendix of \cite{Bierwage10a}). The presence of this propagating component implies that the frequency of the discrete Alfv\'{e}n eigenmode overlaps with the continuous spectrum and does \textit{not} reside inside a gap. Consequently, in regions (IVa) and (V) of figure \ref{fig:scan-a_hirose} and region (IVa) of figure \ref{fig:scan-a_dong}, where $\alpha$TAEs are below the threshold $\alpha_0$ for quasi-marginal stability, continuum damping can be expected to contribute to the threshold for ITG instability.

\begin{figure}[tbp]
\includegraphics[width=1.0\textwidth]
{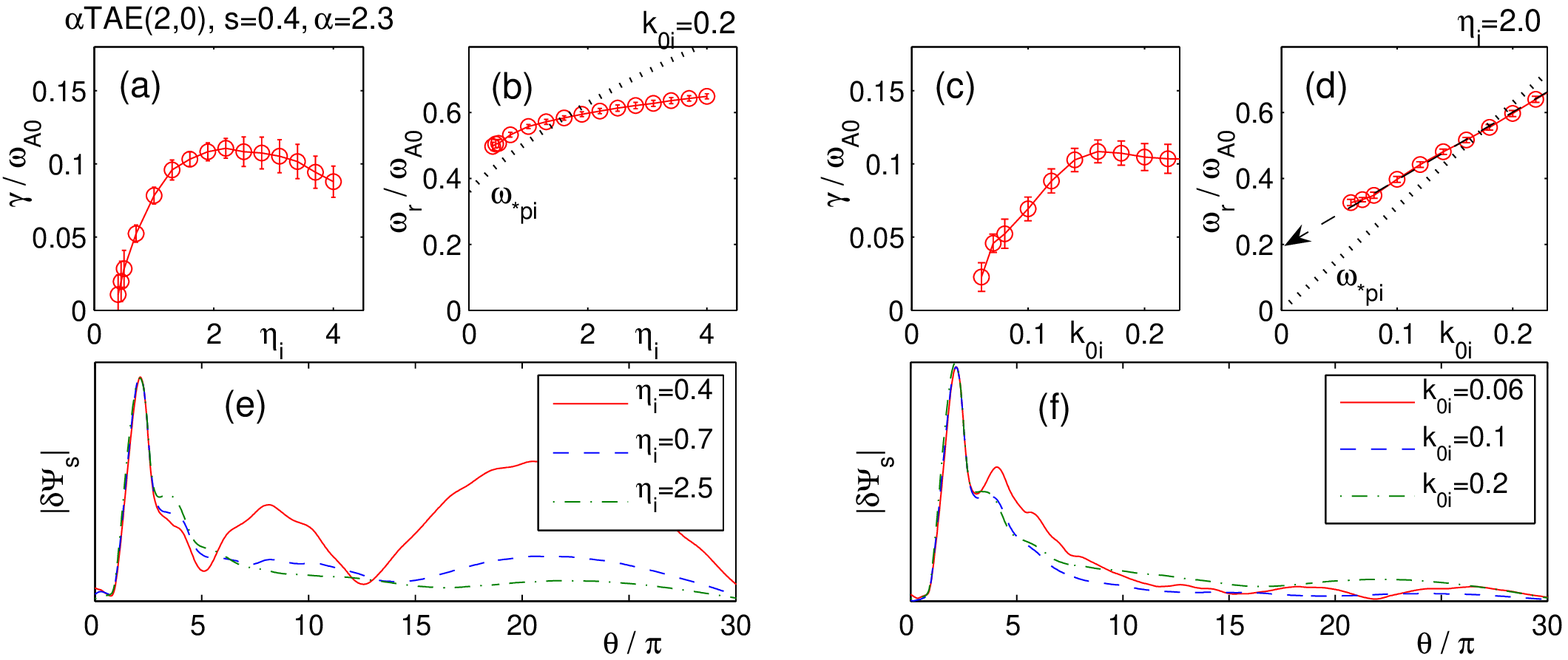}
\caption{Parameter scans with respect to temperature gradient $\eta_{\rm i}$ [(a) and (b)] and wavenumber $\krhoiO$ [(c) and (d)] for an $\alpha$TAE(2,0) in the case with lower shear, $s=0.4$ [region (IVa) in figure \protect\ref{fig:scan-a_hirose}]. Mode structures are shown in (e) and (f). For comparison with the gyrokinetic simulation results (\opencircle), the diamagnetic drift frequency $\omega_{*{\rm pi}}$ (\dotted) is shown in (b) and (d). The arrow in (d) indicates linear extrapolation of $\omega_{\rm r}(\krhoiO)$ towards $\krhoiO\rightarrow 0^+$.}
\label{fig:scaling_hirose}%
\end{figure}

\begin{figure}[tbp]
\includegraphics[width=1.0\textwidth]
{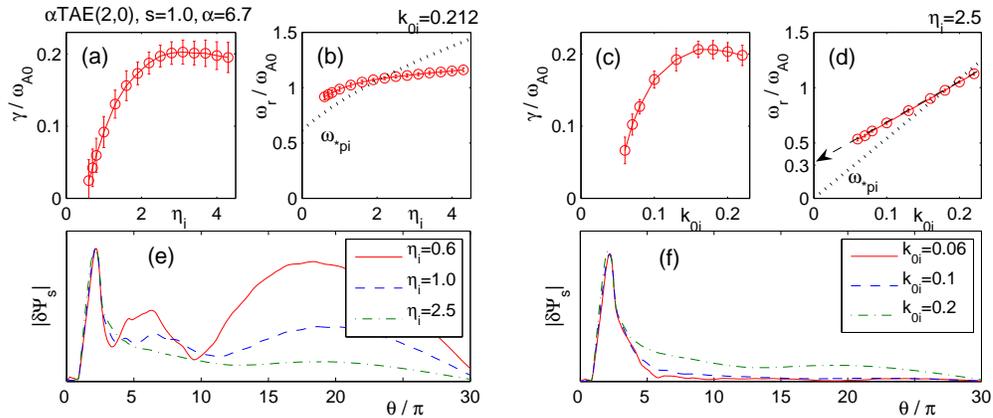}
\caption{Parameter scans with respect to temperature gradient $\eta_{\rm i}$ [(a) and (b)] and wavenumber $\krhoiO$ [(c) and (d)] for an $\alpha$TAE(2,0) in the case with higher shear, $s=1.0$ [region (IVa) in figure \protect\ref{fig:scan-a_dong}]. Mode structures are shown in (e) and (f). For comparison with the gyrokinetic simulation results (\opencircle), the diamagnetic drift frequency $\omega_{*{\rm pi}}$ (\dotted) is shown in (b) and (d). The arrow in (d) indicates linear extrapolation of $\omega_{\rm r}(\krhoiO)$ towards $\krhoiO\rightarrow 0^+$.}
\label{fig:scaling_dong}%
\end{figure}

\subsection{Dependence on temperature gradient $\eta_{\rm i}$ and wavenumber $\krhoiO$}
\label{sec:scans_scaling}

The instabilities observed in the second MHD stable domain in figures~\ref{fig:scan-a_hirose} and \ref{fig:scan-a_dong} all require that the temperature gradient, $\eta_{\rm i}$, and the wavenumber, $\krhoiO$, exceed finite thresholds. In figures \ref{fig:scaling_hirose} and \ref{fig:scaling_dong}, this is demonstrated for modes of the $\alpha$TAE(2,0) branch. In both cases, the thresholds are near $\eta_{\rm i} \sim 0.5$ and $\krhoiO \sim 0.05$ [panels (a) and (c) in both figures]. Since the values of $\alpha$ are chosen to be in region (IVa) of figures \ref{fig:scan-a_hirose} and \ref{fig:scan-a_dong} (i.e., below quasi-marginal stability in the FLR MHD limit, $\alpha < \alpha_0^{(2,0)}$), this threshold can be expected to be due both continuum damping and Landau damping.

\begin{figure}[tbp]
\includegraphics[width=1.0\textwidth]
{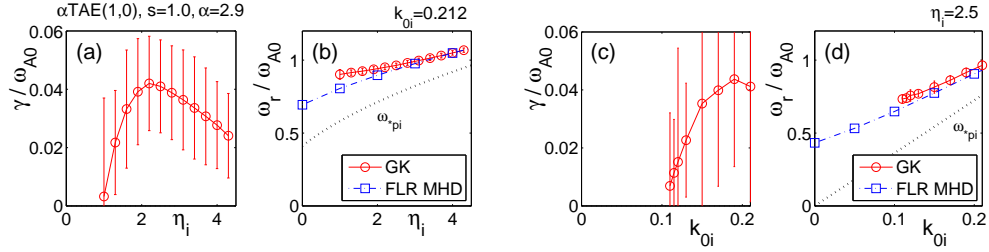}
\caption{Parameter scans with respect to $\eta_{\rm i}$ [(a) and (b)] and $\krhoiO$ [(c) and (d)] for an $\alpha$TAE(1,0) in the case with higher shear, $s=1.0$ [region (III) in figure \protect\ref{fig:scan-a_dong}]. For comparison with the gyrokinetic simulation results (\opencircle), eigenvalues obtained in the FLR MHD limit (\opensquare) \protect\cite{Bierwage10a}, and the diamagnetic drift frequency $\omega_{*{\rm pi}}$ (\dotted) are shown in (b) and (d). Note that the MHD mode is in the quasi-marginally stable regime, so eigenvalues are easily obtained below the threshold for ITG instability using the methods of \protect\cite{Bierwage10a}. The mode structure is shown in figure 3(b) of \protect\cite{Bierwage10a}.}
\label{fig:scaling_atae1}%
\end{figure}

Corresponding results for an $\alpha$TAE(1,0) in the FLR MHD quasi-marginally stable regime [$\alpha > \alpha_0^{(1,0)}$; here, region (III) of figure \ref{fig:scan-a_dong}], are shown in figure \ref{fig:scaling_atae1}. Since continuum damping is negligibly small for this $\alpha$TAE(1,0), the threshold is due to Landau damping only. Nevertheless, the instability thresholds, $\eta_{\rm i} \sim 1.0$ and $\krhoiO \sim 0.1$ [panels (a) and (c)], are higher than for the modes in region (IVa) shown in figures \ref{fig:scaling_hirose} and \ref{fig:scaling_dong}. We conjecture that this is due to the fact that the eigenfrequency $\omega_{\rm r}$ of the $\alpha$TAE$(1,0)$ in figure \ref{fig:scaling_atae1} is significantly larger than $\omega_{*p{\rm i}}$, while $\omega_{\rm r} \approx \omega_{*p{\rm i}}$ for the $\alpha$TAE$(2,0)$ in figures \ref{fig:scaling_hirose} and \ref{fig:scaling_dong}. This implies that the $\alpha$TAE$(1,0)$ in figure \ref{fig:scaling_atae1} is subject to stronger Landau damping [cf.~section \ref{sec:wpi_landau}].

In all three cases shown in figures \ref{fig:scaling_hirose}--\ref{fig:scaling_atae1}, the growth rate $\gamma$ peaks near $\eta_{\rm i} \sim (2\cdots 3)$ and $\krhoi \sim (0.15\cdots 0.2)$. The subsequent decrease in $\gamma$ may be understood by noting that, with increasing $\krhoiO$, large Larmor radii average over the wave field and, with increasing $\eta_{\rm i}$ and constant $\alpha$, the thermal velocity decreases. In both cases, the drive is reduced [cf.~section \ref{sec:wpi_particles}].

Let us now turn our attention to the frequencies in the $\krhoiO$-scans [panel (d) in figures \ref{fig:scaling_hirose}--\ref{fig:scaling_atae1}]. It can be seen that a na\"{i}ve linear extrapolation of $\omega_{\rm r}(\krhoiO)$ to $\krhoiO \rightarrow 0^+$ yields finite frequencies in all cases. In figures \ref{fig:scaling_hirose}(d) and \ref{fig:scaling_dong}(d), this is indicated by arrows; whereas in figure \ref{fig:scaling_atae1}(d), we may follow the data points obtained with the FLR MHD model (\opensquare). Obviously, a part of this frequency is due to the finite frequency of ideal MHD $\alpha$TAEs. In addition, as $\krhoiO$ decreases and the diamagnetic frequency shift becomes smaller, one may expect to observe the effect of coupling to ion sound waves. The associated frequency shift is likely to be responsible for part of the finite frequency obtained for $\krhoiO \rightarrow 0^+$. This conjecture is supported by the analysis presented in section \ref{sec:qforms} below and by the mode structures in figure \ref{fig:scaling_dong}(e) and (f).

In figure \ref{fig:scaling_dong}(e), the amplitude of the continuum wave (large-$|\theta|$ tail) \textit{increases} as $\eta_{\rm i}$ (and, thus, $\gamma$) decreases, which indicates strong continuum damping. In contrast, in figure \ref{fig:scaling_dong}(f), the amplitude of the continuum wave \textit{decreases} as $\krhoiO$ (and both $\gamma$ and $\omega_{\rm r}$) decreases, which indicates that the influence of continuum damping decreases despite the reduced ITG drive. Based on this observation, we conjecture that the mode frequency in \ref{fig:scaling_dong}(d) enters the kinetic thermal ion (KTI) gap \cite{Zonca96, Chen07}, as $\krhoiO$ decreases.

Note that the situation is different in the case with lower shear shown in figure \ref{fig:scaling_hirose}. The mode structures shown in figure \ref{fig:scaling_hirose}(f) indicate a transition from an $\alpha$TAE with dominant (2,0) bound state component to one with dominant (3,0) component, which is subject to stronger continuum damping.

\section{Numerical analysis: SAW equation in quadratic form}
\label{sec:qforms}

In this section, we compute components of the SAW equation in quadratic form and use them to obtain further insight into the $\alpha$-scans in figures \ref{fig:scan-a_hirose} and \ref{fig:scan-a_dong} as well as the $\krhoiO$-scans in figures \ref{fig:scaling_hirose} and \ref{fig:scaling_dong}. For simplicity, the calculations are carried out using the reduced model introduced in section \ref{sec:saw} with $\tauei^T = 0$; i.e., $\delta E_\parallel = 0$. These simplification have no significant qualitative effect and are noticeable here only through minor shifts of the boundaries between regions (III)--(V).

Written in quadratic form and letting $\tauei^T = 0$, equation (\ref{eq:model_saw_red}) becomes
\begin{equation}
i\Phi_{\rm b} = \delta\L, \qquad \delta\L = \delta\L_\omega - \delta W_{\rm f} - \delta W_{\rm ki};
\label{eq:model_saw_dw}
\end{equation}

\noindent where the components of the Lagrangian are
\begin{eqnarray}
\delta\L_\omega &= -\frac{1}{N}\int_{-\theta_{\rm max}}^{\theta_{\rm max}} {\rm d}\theta\, \left(V_{{\rm m},\omega} + V_{{\rm m},\kappa} \right)\left|\delta\Psi_{\rm s}\right|^2,
\label{eq:model_saw_dw_flr}
\\
\delta W_{\rm f} &= \frac{1}{N}\int_{-\theta_{\rm max}}^{\theta_{\rm max}} {\rm d}\theta\, \left[ \left|\delta\Psi_{\rm s}'\right|^2 + V\left|\delta\Psi_{\rm s}\right|^2\right],
\label{eq:model_saw_dw_dwf}
\\
\delta W_{\rm ki} &= \frac{1}{N}\int_{-\theta_{\rm max}}^{\theta_{\rm max}} {\rm d}\theta\, \delta\Psi_{\rm s}^* V_{\rm ki}\delta\Psi_{\rm s},
\label{eq:model_saw_dw_kic}
\end{eqnarray}

\noindent for a computational domain $-\theta_{\rm max} \leq \theta \leq \theta_{\rm max}$. The normalization constant is given by $N = \int_{-\theta_{\rm max}}^{\theta_{\rm max}} {\rm d}\theta\, |\delta\Psi_{\rm s}|^2/f$. This choice is motivated by the discussion in the Appendix of \cite{Bierwage10a} and is crucial to make parameter scans meaningful. \textsc{awecs} imposes fixed boundary conditions, $\psi(\pm\theta_{\rm max}) = 0$, buffered by a region which is $0.2\times\theta_{\rm max}$ wide and where artificial damping is applied. Thus, the energy flux through the boundary is zero: $\Phi_{\rm b} = iN^{-1} [\delta\Psi_{\rm s}^* \delta\Psi_{\rm s}']_{-\theta_{\rm max}}^{\theta_{\rm max}} = 0$.

$\delta\L_\omega$ contains the $\omega$-dependent contributions; namely, the inertia term (with FLR) and the FLR correction of the ``MPC+MFC'' term. The magnitude of the ideal MHD potential energy, $\delta W_{\rm f}$, measures the strength of FLB versus ideal MHD ballooning-interchange drive (the standard definition is $\delta\hat{W}_{\rm f} = (N/2)\delta W_{\rm f}$). The complex number $\delta W_{\rm ki}$ measures the effect of kinetic thermal ion compression, both non-resonant and resonant contributions.

The imaginary part of (\ref{eq:model_saw_dw}) contains no information which is of interest here [except that $\gamma \propto {\rm Im}(\delta W_{\rm ki})$ in the FLR MHD stable domains], so it suffices to analyze its real part, which we write as
\begin{equation}
E_\omega - \delta W_{\rm f} - {\rm Re}(\delta W_{\rm ki}) = C_0\gamma^2;
\label{eq:disp_r}
\end{equation}

\noindent where the frequency-dependent term is decomposed as ${\rm Re}(\delta\L_\omega) = E_\omega - C_0\gamma^2$, and $E_\omega$ has the form
\begin{equation}
\fl E_\omega = C_0(\Omega^2 - \Omega_0^2), \quad
\Omega = \omega_{\rm r} - \Omega_1, \quad
\Omega_0^2 = \Omega_1^2 + \Omega_2^2.
\end{equation}

\noindent The terms $\Omega_1$ and $\Omega_2$ are defined as follows:
\begin{eqnarray}
\fl {\rm Re}(\delta\L_\omega) = & -\omega_{\rm r} \underbrace{\left<\omega_{*{\rm i}} (1 - \Gamma_0\Upsilon_1)/b_{\rm 0i} - 2\Omega_\kappa(1 - \Gamma_0\Delta_1)/b_{\rm 0i}\right>_\Psi}_{2 C_0 \Omega_1}
\\
\fl & - \underbrace{\left<2\Omega_\kappa\omega_{*p{\rm i}} [1 - \Gamma_0\Upsilon_{2\kappa}/(1 + \eta_{\rm i})]/b_{\rm 0i}\right>_\Psi}_{C_0 \Omega_2^2}
+ (\omega_{\rm r}^2 - \gamma^2)\underbrace{\left<(1 - \Gamma_0)/b_{\rm i}\right>_\Psi}_{C_0}; \nonumber
\end{eqnarray}

\noindent where $\left<X\right>_\Psi \equiv N^{-1} \int_{-\theta_{\rm max}}^{\theta_{\rm max}}{\rm d}\theta\, X|\delta\Psi_{\rm s}|^2$, and the functions $\Gamma_0$, $\Delta_1$, $\Upsilon_1$ and $\Upsilon_{2\kappa}$ are defined in equation (13) of \cite{Bierwage10a}.

\begin{figure}[tbp]
\includegraphics[width=1.0\textwidth]
{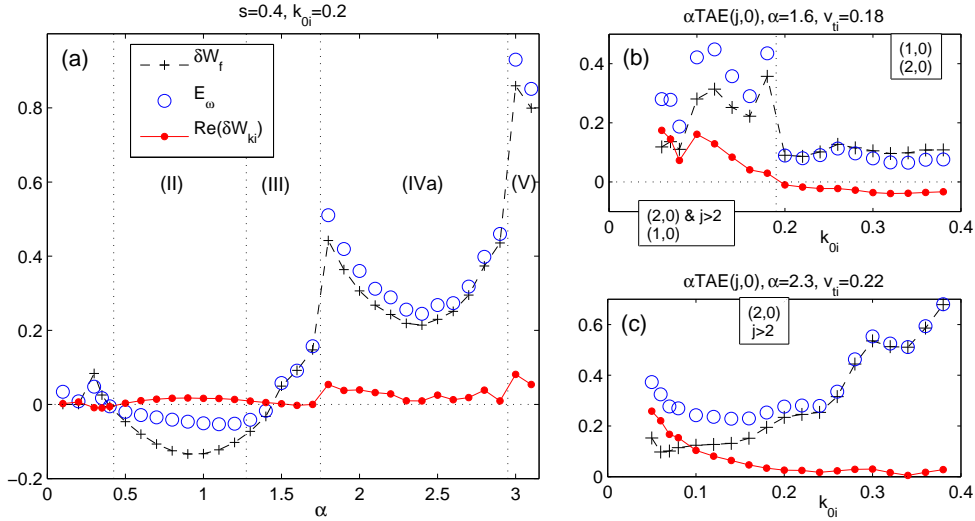}
\caption{Analysis of quadratic forms $\delta W_{\rm f}$ (+), $E_\omega$ (\opencircle) and ${\rm Re}(\delta W_{\rm ki})$ (\fullcircle) in the case with $s=0.4$. In (a), the $\alpha$-dependence is shown, with regions (II)--(V) corresponding to those in figure \protect\ref{fig:scan-a_hirose}. In (b) and (c), $\krhoiO$-dependence is shown for $\alpha=1.6$ in region (III) and $\alpha=2.3$ in region (IVa), respectively. The inset boxes with labels $(j,p)$ indicate the main bound state components, starting with the dominant one.}
\label{fig:scan-a_qforms_hirose}%
\end{figure}

\begin{figure}[tbp]
\includegraphics[width=1.0\textwidth]
{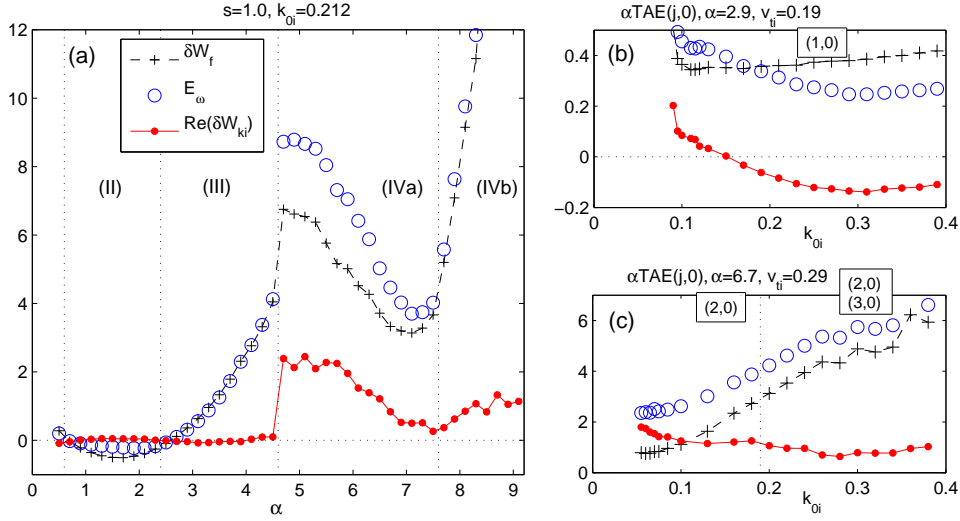}
\caption{Analysis of quadratic forms $\delta W_{\rm f}$ (+), $E_\omega$ (\opencircle) and ${\rm Re}(\delta W_{\rm ki})$ (\fullcircle) in the case with $s=1.0$. In (a), the $\alpha$-dependence is shown, with regions (II)--(IV) corresponding to those in figure \protect\ref{fig:scan-a_dong}. In (b) and (c), $\krhoiO$-dependence is shown for $\alpha=2.9$ in region (III) and $\alpha=6.7$ in region (IVa), respectively. The inset boxes with labels $(j,p)$ indicate the main bound state components, starting with the dominant one.}
\label{fig:scan-a_qforms_dong}%
\end{figure}

The results for the case with $s=0.4$ are shown in figure \ref{fig:scan-a_qforms_hirose} and those for $s=1.0$ in figure \ref{fig:scan-a_qforms_dong}. Note that each point in figures \ref{fig:scan-a_qforms_hirose} and \ref{fig:scan-a_qforms_dong} represents a snapshots taken at a given time $t$. In order to reduce the level of phase-space discretization noise, we use a larger number of markers here: $N_{\rm m} = 8192\times 13$.

The discussion in section \ref{sec:qforms_rflb} focuses on $\delta W_{\rm f}$, and section \ref{sec:qforms_kic_vs_rflb} on ${\rm Re}(\delta W_{\rm ki})$. $E_\omega$ is shown for completeness and $C_0 \gamma^2$ is omitted because it follows from the sum of the other three terms [and it is very small, except in the FLR MHD ballooning unstable domain, region (II)]. In section \ref{sec:qforms_compound}, we inspect the mode structures with respect to coupling between different bound state components.

\subsection{Field line bending versus ballooning-interchange drive}
\label{sec:qforms_rflb}

The qualitative form of the $\alpha$-dependence of the ideal MHD potential energy, $\delta W_{\rm f}(\alpha)$, is similar in both figures \ref{fig:scan-a_qforms_hirose}(a) and \ref{fig:scan-a_qforms_dong}(a). For each $\alpha$TAE branch, $\delta W_{\rm f}(\alpha)$ has a parabola-like shape and the intersections of these parabolas coincide with the points where the dominant instability switches branches. The minima in $\delta W_{\rm f}$ correspond to the points where the growth rates are largest.

The minimum of the parabola traversing region (II) is negative, $\delta W_{\rm f} < 0$. In this region, ideal MHD ballooning drive is stronger than field line bending and the $\alpha$TAE$(1,0)$ branch is replaced by unstable ideal MHD ballooning modes, modified by FLR and kinetic effects. The points where FLB and ballooning drive balance, $\delta W_{\rm f} = 0$, coincide with the ideal MHD ballooning stability boundaries: $\alpha_{\rm crit,1} \approx 0.4$ and $\alpha_{\rm crit,2} \approx 1.5$ in figure \ref{fig:scan-a_qforms_hirose}(a), and $\alpha_{\rm crit,1} \approx 0.6$ and $\alpha_{\rm crit,2} \approx 2.6$ in figure \ref{fig:scan-a_qforms_dong}(a). This implies that FLR and kinetic effects do not modify the ideal MHD mode structures, only the eigenvalues. In the low-$\alpha$ part of region (III), $\delta W_{\rm f}$ is still negative, but small enough to enable FLR effects to stabilize ideal MHD ballooning modes.

\subsection{Kinetic compression versus ideal MHD potential energy}
\label{sec:qforms_kic_vs_rflb}

In region (II) of both figures \ref{fig:scan-a_qforms_hirose}(a) and \ref{fig:scan-a_qforms_dong}(a), we find ${\rm Re}(\delta W_{\rm ki}) > 0$. This confirms that the up-shift in the frequencies seen in figures \ref{fig:scan-a_hirose}(b) and \ref{fig:scan-a_dong}(b), where we compared gyrokinetic simulation results (\opencircle) with FLR MHD results ($\omega_{\rm BM}$, \chain), is due to kinetic compression.

In region (III), as $\alpha$ is increased well beyond the second MHD stability boundary, ${\rm Re}(\delta W_{\rm ki})$ becomes small compared to $\delta W_{\rm f}$ in both figures \ref{fig:scan-a_qforms_hirose}(a) and \ref{fig:scan-a_qforms_dong}(a). It can even be seen to drop below zero; especially, when $\krhoiO$ is increased as shown in figures \ref{fig:scan-a_qforms_hirose}(b) and \ref{fig:scan-a_qforms_dong}(b), which indicates that the resonant contribution dominates over the non-resonant contribution. Thus, the eigenmodes in region (III) tend to have an ``incompressible'' response, in the sense that coupling to the ion sound branch is small; especially, for wavenumbers $\krhoiO$ well above ITG threshold. This is consistent with figures \ref{fig:scan-a_hirose}(b) and \ref{fig:scan-a_dong}(b), where we saw that in region (III), the frequency from the gyrokinetic simulation (\opencircle) closely matches that of the underlying FLR MHD mode ($\omega_{\alpha{\rm TAE}}^{(1,0)}$, \chain).

In region (IVa), we find ${\rm Re}(\delta W_{\rm ki}) \gtrsim 0$. The ratio ${\rm Re}(\delta W_{\rm ki})/\delta W_{\rm f}$ reaches 10\% in figures \ref{fig:scan-a_qforms_hirose}(a) and 50\% in figures \ref{fig:scan-a_qforms_dong}(a), which implies that kinetic compression causes a significant up-shift in the frequency. The $\krhoiO$-scans in figures \ref{fig:scan-a_qforms_hirose}(b) and \ref{fig:scan-a_qforms_dong}(b) show that, for sufficiently small wavenumbers $\krhoiO$, ${\rm Re}(\delta W_{\rm ki})$ tends to rise above $\delta W_{\rm f}$. This confirms our earlier conjecture that non-resonant kinetic compression is responsible for the finite frequency obtained in figures \ref{fig:scaling_hirose} and \ref{fig:scaling_dong} by na\"{i}ve extrapolation towards $\krhoiO \rightarrow 0^+$.

Region (V) in figure \ref{fig:scan-a_qforms_hirose}(a) is similar to region (IVa), and region (IVb) in figure \ref{fig:scan-a_qforms_dong}(a) is similar to region (III).

\subsection{Coupling of multiple bound states}
\label{sec:qforms_compound}

In the $\alpha$-scans analyzed in figures \ref{fig:scan-a_qforms_hirose}(a) and \ref{fig:scan-a_qforms_dong}(a), switching between different branches of $\alpha$TAEs is observed at the boundaries of regions (III)--(V). As pointed out above, this is correlated with minimization of $\delta W_{\rm f}$ (cf.~section \ref{sec:qforms_rflb}), with possible small contributions from continuum damping and wave-particle interaction. In this section, we inspect the transitions seen in the $\krhoiO$-scans in figures \ref{fig:scan-a_qforms_hirose}(b) and \ref{fig:scan-a_qforms_dong}(c). The mode structures are not shown explicitly; instead panels (b) and (c) in figures \ref{fig:scan-a_qforms_hirose} and \ref{fig:scan-a_qforms_dong} have inset boxes, where we list the types of bound states $(j,p)$ seen in the corresponding mode structures, starting with the dominant component.

In figure \ref{fig:scan-a_qforms_hirose}(b), a transition occurs around $\krhoiO \approx 0.19$, where ${\rm Re}(\delta W_{\rm ki})$ switches sign. For $\krhoiO < 0.19$, where ${\rm Re}(\delta W_{\rm ki}) > 0$, the modes excited have complex mode structures which consist of a dominant $\alpha$TAE$(2,0)$ component and also strong components with $j>2$. In addition, the mode structure exhibits a clear but smaller peak in the central potential well; i.e., an $\alpha$TAE$(1,0)$ component. For $\krhoiO > 0.19$, where ${\rm Re}(\delta W_{\rm ki}) < 0$, the $\alpha$TAE$(1,0)$ component is dominant. The slight increase in $\delta W_{\rm f}$ around $\krhoiO \approx 0.26$ coincides with the location where the primary $\alpha{\rm TAE}(1,0)$ couples to a $(2,0)$ component.

In the case of figure \ref{fig:scan-a_qforms_hirose}(c), the non-ideal $\alpha$TAE$(2,0)$ is the dominant instability in the entire $\krhoiO$ range scanned, but the mode structure also contains minor components with $j>2$.

The mode structures are less complex in the case with higher shear, $s=1.0$, shown in figure \ref{fig:scan-a_qforms_dong}(b) and (c). Here, fewer bound state components are coupled together, and an $\alpha$TAE$(1,0)$ and an $\alpha$TAE$(2,0)$ can be clearly identified, respectively. For $\krhoiO \gtrsim 0.2$ in figure \ref{fig:scan-a_qforms_dong}(c), the $\alpha$TAE$(2,0)$ acquires a minor $(3,0)$ component.

\section{Numerical analysis: wave-particle resonance}
\label{sec:wpi}

In this section, we examine in detail the wave-particle interactions responsible for the excitation of the Alfv\'{e}nic instabilities described in sections \ref{sec:atae_comp} and \ref{sec:qforms} by analyzing the properties of the relevant driving term for Alfv\'{e}nic ITG (AITG) modes \cite{Zonca99, Dong99},
\begin{equation}
\left<\omegad J_0 \delta G\right>_{\rm i} \approx \sum_j (\Delta v)_j^3 \omega_{{\rm d}j} J_0(\lambda_j) \delta G_j \equiv \sum_j (\Delta H)_j;
\label{eq:def_dh}
\end{equation}

\noindent where the subscript $j$ labels a phase space marker. We use the normalization
\begin{equation}
\hat{\enr} = \enr/\Ti, \quad \hat{v}_\perp = v_\perp/\vti, \quad \hat{v}_\parallel = v_\parallel/\vti, \quad \hat{\omega} = \omega/\omAO;
\end{equation}

\noindent where the hats are omitted in the following.

The method of analyzing contributions from individual markers allows to obtain much more detailed insight than an inspection of the imaginary component of the quadratic forms in equation (\ref{eq:model_saw_dw}) could give. In section \ref{sec:wpi_landau}, a method to eliminate Landau damping and measure the associated damping rate is described and used. In section \ref{sec:wpi_particles}, we analyze the dependence of $|\Delta H|_j$ on $\enr$, $v_\perp$ and $v_\parallel$. The characteristic energy-dependence expected for AITG instabilities is reproduced numerically and the instability drive is confirmed to be due to interactions with the high-energy tail of the equilibrium distribution \cite{Zheng00, Zonca00}.

\subsection{Landau resonance}
\label{sec:wpi_landau}

The quantity $Q_{\rm i} = (\omega_{*{\rm i}}^T - \omega)/\Ti$ in equation (\ref{eq:model_vlasov_gke2}) originates from the phase-space gradient operator, $\hat{Q}_s = \omega\partial_\enr + (qR_0)^2\omLs^{-1}({\bm k}_\perp\times\hat{\bm b})\cdot\nablab$, acting on the Maxwellian distribution $\feqi$. For modes satisfying $\gamma \ll \omega_{\rm r}$, the boundary in velocity space at which the direction of Landau damping is inverted is given by the condition $Q_{\rm i} = 0$, which can be expressed as a condition on the energy as
\begin{equation}
\enr_{\rm crit} = \frac{3}{2} + \frac{1}{\eta_{\rm i}} \frac{\omega_{\rm r} - \omega_{*{\rm i}}}{\omega_{*{\rm i}}}, \qquad (\gamma \ll \omega_{\rm r});
\label{eq:ecrit}
\end{equation}

\noindent where the normalized diamagnetic frequency is given by equation~(\ref{eq:norm_omgf_omgk}). Particles with energies $\enr > \enr_{\rm crit}$ are capable of driving an instability via inverse Landau damping.

It is possible to eliminate the effect of Landau damping by dropping the contributions $(\Delta H)_j$ of particles with energies $\enr < \enr_{\rm crit}$ if $\omega_{\rm r} \gg \gamma$ and if at least one of the following conditions is satisfied:
\begin{itemize}
\item[(a)]  the frequency shift caused by kinetic compression [${\rm Re}(\delta W_{\rm ki})$ in (\ref{eq:disp_r})] is much smaller than the mode frequency $\omega_{\rm r}$; and/or
\item[(b)]  the particles with energies $\enr < \enr_{\rm crit}$ contribute only a negligible part of that frequency shift.
\end{itemize}

\noindent Assuming that this is the case (checked \textit{a posteriori}), we let $(\Delta H)_j = 0$ for $\enr_j < \enr_{\rm crit}$ in equation (\ref{eq:def_dh}). The results obtained from such a screening test are summarized in table~\ref{tab:screen} for three values of $\alpha$, sampling regions (I), (III) and (IVa) in figure~\ref{fig:scan-a_dong}.

In all cases shown in table~\ref{tab:screen}, the frequencies change only by a small amount ($\sim 3\%$); presumably, because at least condition (a) is satisfied, since, for the present parameters, the diamagnetic frequency shift is much larger than that caused by kinetic compression. Hence, the value of $\enr_{\rm crit}$ and the resonance condition are not affected by the screening test and the procedure is meaningful. Consequently, the difference in the growth rates shown in table~\ref{tab:screen} gives a measurement of the Landau damping rate, $-\gamma_{\rm LD}$, which, for the three cases inspected, lies in the range $0.006 \lesssim -\gamma_{\rm LD} \lesssim 0.010$. This is somewhat lower than or comparable to the continuum damping rate found for $\alpha$TAEs in region (IV) of figures \ref{fig:scan-a_hirose} and \ref{fig:scan-a_dong} (see figure 6 in \cite{Bierwage10a}).

\begin{table}
\caption{\label{tab:screen}Growth rate $\gamma$ and frequency $\omega_{\rm r}$ of ITG-driven Alfv\'{e}nic instabilities in regions (I), (III) and (IVa) in figure \protect\ref{fig:scan-a_dong}. Results for the case where all particles are included (``all $\enr$'') are compared to the case where only contributions from particles with energies satisfying $\enr \geq \enr_{\rm crit}$ are included (``$\enr_{\rm high}$''). $\enr$ is normalized by $\vti^2$, and $\omega_{\rm r}$ and $\gamma$ by $\omAO$.}
\begin{indented}
\item[]\begin{tabular}{@{}rccccccccc}
\br
& \centre{2}{region (I)} & \multicolumn{2}{c}{region (III)} & \multicolumn{2}{c}{region (IVa)} \\
& \centre{2}{$\alpha = 0.5$, $\enr_{\rm crit} = 2.4$} & \multicolumn{2}{c}{$\alpha = 2.9$, $\enr_{\rm crit} = 2.9$} & \multicolumn{2}{c}{$\alpha = 6.0$, $\enr_{\rm crit} = 2.7$} \\
\ns & \crule{2} & \crule{2} & \crule{2} \\
& all $\enr$ & $\enr_{\rm high}$ & all $\enr$ & $\enr_{\rm high}$ & all $\enr$ & $\enr_{\rm high}$ \\
\mr
$\gamma$: & 0.016 & 0.024 & 0.034 & 0.044 & 0.092 & 0.098 \\
$\omega_{\rm r}$: & 0.282 & 0.277 & 0.961 & 0.935 & 1.222 & 1.191 \\
\br
\end{tabular}
\end{indented}
\end{table}

\subsection{Contributions from individual particles}
\label{sec:wpi_particles}

\begin{figure}[tbp]
\includegraphics[width=1.0\textwidth]
{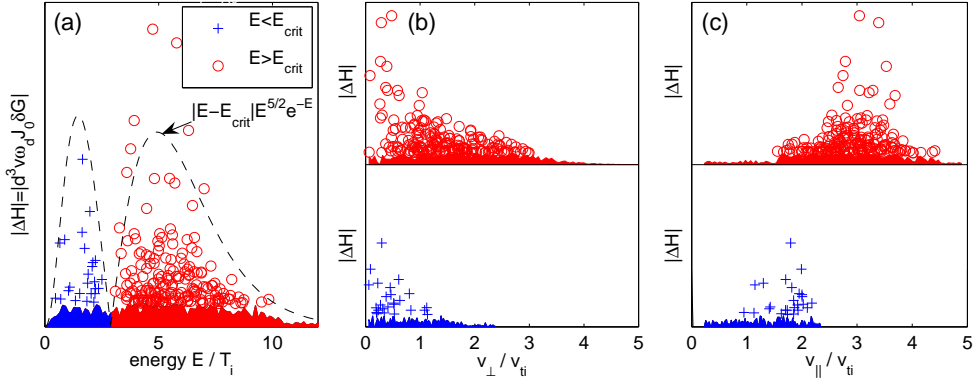}
\caption{Contribution of individual marker particles to the kinetic compression term $\left<\omegad J_0 \delta G\right>_{\rm i} = \sum_j(\Delta H)_j$ in equation (\protect\ref{eq:model_maxw_vort1}). Results are shown for $\alpha = 2.9$ in region (III) of figure~\protect\ref{fig:scan-a_dong}, where an $\alpha$TAE(1,0) is the dominant instability. A snapshot of the quantity $|\Delta H|_j$ is plotted for each marker particle $j$ as a function of (a) the particle energy $\enr = (v_\perp^2 + v_\parallel^2)/2$, (b) the perpendicular velocity $v_\perp$, and (c) the parallel velocity $v_\parallel$, where all velocities are normalized by $\vti$. The marker particles are separated into two populations: $\enr > \enr_{\rm crit} \approx 2.9$ ($\opencircle$) and $\enr < \enr_{\rm crit}$ ($+$), with $\enr_{\rm crit}$ given by equation (\ref{eq:ecrit}). In addition, (a) shows the expected energy scaling $|\enr - \enr_{\rm crit}|\enr^{5/2}\exp(-\enr)$ (\broken).}
\label{fig:wpi}%
\end{figure}

The contributions of individual particles to the kinetic compression term $\left<\omegad J_0 \delta G\right>_{\rm i}$ is shown in figure~\ref{fig:wpi}, where the quantity $|\Delta H|_j$ is plotted as a function of (a) the particle energy $\enr$, (b) the perpendicular velocity $v_\perp$, and (c) the parallel velocity $v_\parallel$. For this analysis, we have chosen an $\alpha$TAE(1,0) from region (III) in figure~\ref{fig:scan-a_dong} at $\alpha = 2.9$. Similar results are obtained for AITG instabilities in regions (I), (IV) and (V).

The plots constitute snapshots taken at a certain time $t$. Note that the marker particles are loaded uniformly in energy $\enr$, so the density of markers in the plots does not reflect the number density of the Maxwellian-distributed physical particles.

\paragraph{Energy dependence.} Equations (\ref{eq:model_vlasov_gke2}) and (\ref{eq:def_dh}) imply that the energy dependence of $|\Delta H|(\enr)$ should follow the scaling $|\enr^{1/2} Q_{\rm i} \feqi \omegadi \delta S_{1{\rm i}}| \propto |\enr - \enr_{\rm crit}|\enr^{5/2}\exp(-\enr)$ which, indeed, gives an accurate envelope (\dashed) for the data points in figure~\ref{fig:wpi}(a). The location of the minimum at $\enr \sim 2.9$ corresponds to the critical energy $\enr_{\rm crit}$ given by equation (\ref{eq:ecrit}), so the instability drive is due to particles with $\enr > 2.9$, while Landau damping is exerted by particles with $\enr < 2.9$.

\paragraph{Dependence on $v_\perp$.} Figure~\ref{fig:wpi}(b) shows that particles with lower $v_\perp$ tend to have larger contributions. The peaks shift towards lower $v_\perp$ when $\krhoiO$ is increased (not shown). This behavior reflects the fact that maximal wave-particle interactions require an optimal ration of Larmor radii to perpendicular wavelength. In the case of figure~\ref{fig:wpi}(b), we have $\krhoiO = 0.2$ and the strongest interactions occur around $v_\perp \sim 1$. Given that $\alpha = 2.9$, $s=1.0$, and the mode peak is localized around $\theta \in [-\pi,\pi]$, we have $\sqrt{f} \sim 3$. Thus, Larmor radii satisfying $\lambda_{\rm i} = k_\perp v_\perp/\omLiO = \sqrt{f} \krhoiO v_\perp \sim 0.6$ are favored.

\paragraph{Dependence on $v_\parallel$.} Figure~\ref{fig:wpi}(c) shows that particles with larger $v_\parallel$ tend to have larger contributions. The strongest response occurs around $v_\parallel \sim 3$. One reason is that the energies required for the instability drive through $|\Delta H|_j$ are relatively large [$\enr \sim 5$; cf.~figure~\ref{fig:wpi}(a)], so that large $v_\parallel$ compensate for the low values of $v_\perp$ [cf.~figure~\ref{fig:wpi}(b)] dictated by the constraint $\lambda_{\rm i} < 1$. The actual shape of the envelope for the data points $|\Delta H|_j(v_\parallel)$ in figure~\ref{fig:wpi}(c), is determined by the resonance conditions for passing ions, which has the form,
\begin{equation}
0 = v_\parallel[k_\parallel - p/(q R_0)] - \omega \qquad {\rm (not\; normalized)};
\label{eq:resonance}
\end{equation}

\noindent where $p = 0,\pm 1, ...$ are the side-bands generated by toroidal curvature.

For example, for the $\alpha$TAE$(1,0)$ used in figure~\ref{fig:wpi}, the $k_\parallel$ spectrum is roughly Lorentzian-shaped, centered around zero and with a half-width around $1.5/(qR_0)$ (cf.~figure 4 in \cite{Joiner08}), the thermal transit frequency is $\omega_{\rm ti} \approx 0.14$ [equation (\ref{eq:norm_omgf_omgk})], and the mode frequency is $\omega_{\rm r} \approx 0.9 \gg \gamma \approx 0.05$ (figure \ref{fig:scan-a_dong}). Thus, for particles with $v_\parallel \sim 3$, which show maximal response on the driving side of the distribution ($\enr > \enr_{\rm crit}$), equation (\ref{eq:resonance}) becomes $0.9 \approx 3(\hat{k}_\parallel - p\times 0.14)$, with $\hat{k}_\parallel = q R_0 k_\parallel$. Hence, for $p=0$ these particles interact primarily with the $\hat{k}_\parallel = 0.3$ component of the wavenumber spectrum. Interactions with components $\hat{k}_\parallel \gtrsim 1$ may be expected to be ineffective since they require high-order resonances, $p\geq 5$.

\section{Discussion}
\label{sec:discuss}

The results of the present study may be understood within a theoretical framework which uses a variational formulation based on a variational principle and the assumption that the eigenfunctions have a two-scale structure. The scale separation allows to write the Lagrangian in the form of a so-called generalized fishbone-like dispersion relation (GFLDR), which follows from asymptotic matching between the solution for the ``ideal region'' ($|\theta| \lesssim 1$) and that for the ``inertial layer'' ($|\theta| \gg 1$) (for a review, see \cite{Chen07}). In our specific case, this theory unifies the concepts of kinetic ballooning modes (KBM), beta-induced Alfv\'{e}n eigenmodes (BAE) and Alfv\'{e}nic ITG (AITG) modes, each of which captures the role of a particular physical mechanism.

$\alpha$TAEs cannot be described by a GFLDR because there is no characteristic separation of scales in the mode structures. The eigenvalue is determined in the region between the turning points of the potential well in which the eigenmode is trapped, so all relevant physical processes act in the same narrow region of size $\Delta\theta \approx 2\pi$. However, the physical mechanisms themselves are the same as in those cases where a two-scale structure exists, so the conceptual framework which was constructed from the analysis of the GFLDR is useful to explain the properties of $\alpha$TAEs in the presence of FLR effects and kinetic compression.

Let us write the SAW equation in quadratic form (\ref{eq:model_saw_dw}) as
\begin{equation}
-i\Phi_{\rm b} + \delta\L_\omega - \delta W_{\rm ki} = \delta W_{\rm f}
\quad \stackrel{\rm GFLDR}{\longrightarrow} \quad
i\Lambda = \delta W_{\rm f}.
\label{eq:disp}
\end{equation}

\noindent As indicated in equation (\ref{eq:disp}), the term $i\Lambda(\omega)$, which describes the physics of the  ``inertial layer'' in the standard GFLDR, includes the following contributions:
\begin{enumerate}
\item[(i)]  The boundary term $\Phi_{\rm b}$ accounts for continuum damping (i.e., resonant absorption and phase mixing), which occurs whenever the frequency of an eigenmode overlaps with the continuous spectrum in the region where the mode amplitude is finite.
\item[(ii)]  The ideal-MHD limit of $\delta\L_\omega \stackrel{\rm ideal}{\longrightarrow} \omega^2$ yields ideal MHD inertia.
\item[(iii)]  $\delta\L_\omega$ also contains FLR corrections associated with the inertia term (i.e., diamagnetic drift) and pressure-curvature coupling (i.e., FLR ballooning). These FLR effects (``KBM physics'') were the subject of investigation in the companion paper \cite{Bierwage10a}. Due to FLR effects, the contribution of ion inertia decays with increasing $\theta$ (decreasing radial scale length), so the Alfv\'{e}n velocity diverges until electron inertia and/or collisions (not included here) take over.
\end{enumerate}

\noindent Kinetic thermal ion compression effects, captured by $\delta W_{\rm ki}$, are separated into two categories: non-resonant and resonant:
\begin{enumerate}
\item[(iv)]  The \textit{non-resonant} contribution of kinetic compression (``BAE physics'') accounts for coupling between the shear Alfv\'{e}n wave branch and the ion sound wave branch via geodesic curvature.
\item[(v)]  The \textit{resonant} contribution of kinetic compression represents (inverse) Landau damping in the presence of a temperature gradient (here, ion temperature gradient, ``AITG physics'').
\end{enumerate}

\noindent Together with the ideal MHD potential energy, $\delta W_{\rm f}$ (which includes the effects of geometry, field line bending, and free expansion energy associated with the pressure gradient), the boundary term $\Phi_{\rm b}$ (i) and the inertia term $\omega^2$ (ii) describe ideal MHD shear Alfv\'{e}n waves and eigenmodes. The non-ideal effects (iii)--(v) influence an ideal MHD eigenmode by altering its frequency, growth rate and mode structure. In addition, they introduce gaps and, hence, various accumulation points in the continuous SAW spectrum, which serve as birth points for various types of discrete Alfv\'{e}n eigenmodes \cite{Chen07}.

\paragraph{Normal modes.} Here and in the companion paper \cite{Bierwage10a}, we have seen these mechanisms at work on two types of ideal MHD eigenmodes: ballooning modes and $\alpha$TAEs. Away from the ideal MHD marginal stability boundaries, these branches were clearly identified by comparison with FLR MHD shooting code results from \cite{Bierwage10a} (see figures \ref{fig:scan-a_hirose} and \ref{fig:scan-a_dong} above). Near the ideal MHD marginal stability boundaries, all terms in equation (\ref{eq:disp}) tend to be small and comparable, so different physics and different types of eigenmodes are mixed.

As we scan through the second MHD ballooning stable domain by increasing the normalized pressure gradient $\alpha$ for constant magnetic shear $s$, different branches of $\alpha$TAEs are excited in turns. The switching between branches is found to be correlated with the minimization of ideal MHD potential energy, $\delta W_{\rm f}$ [see figures \ref{fig:scan-a_qforms_hirose}(a) and \ref{fig:scan-a_qforms_dong}(a)].

\paragraph{Resonant drive with ITG.} The broadening of the ideally unstable domain due to wave-particle resonances is seen as predicted theoretically \cite{Zonca96, Zonca99, Cheng82} and previously shown numerically in local \cite{Hirose94, Dong04, Dong99, Zhao02} and global \cite{Falchetto03, Mishchenko08} simulations. Measurements such as those in figure~\ref{fig:wpi}(c) illustrate clearly that, the maximal drive is due to particles in the range $2 \lesssim v_\parallel/\vti \lesssim 4$. This is consistent with the analytic estimate that can be made by maximizing the strength of wave-particle interactions as discussed in \cite{Zheng00, Zonca00}. Thus, in order to correctly estimate the importance of the transit resonance (e.g., when deriving model equations using formal ordering arguments), the reference velocity should be chosen several times larger than the thermal velocity $\vti$.

The analysis in \cite{Zonca99, Zonca98} led to the conclusion that ITG-drive is effective only when an eigenmode is not far from ideal MHD marginal stability. This statement, which is based on an analysis of the GFLDR (two-scale assumption) in the first MHD ballooning stable domain, should be rendered more precisely in the light of our numerical simulations, which do not impose the scale-separation constraint and show a destabilization of $\alpha$TAEs in the second MHD stable domain, far away from marginal stability. What one can say in general is that eigenmodes which minimize $\delta W_{\rm f}$ are most easily destabilized via ITG-drive. This can be seen by comparing figures \ref{fig:scan-a_hirose} and \ref{fig:scan-a_dong} with figures \ref{fig:scan-a_qforms_hirose}(a) and \ref{fig:scan-a_qforms_dong}(a).

Consistently with many earlier works, the resonant destabilization via ITG is found to require a finite temperature gradient, $\eta_{\rm i} \gtrsim \O(0.5)$, and normalized wavenumbers of the order $\krhoiO = k_\vartheta\vti/\omLiO \gtrsim \O(0.1)$ (cf.~section \ref{sec:scans_scaling}). The finite instability thresholds are due to resonant wave absorption by both thermal ions (Landau damping, $\gamma_{\rm LD}$) and shear Alfv\'{e}n continuous spectrum (continuum damping, $\gamma_{\rm CD}$). Our numerical analyses in section~\ref{sec:wpi_landau} show that both absorption rates can be comparable: $-\gamma_{\rm LD} \lesssim -\gamma_{\rm CD} \sim \O(10^{-2})$. In particular, this is true in regimes where the dominant instabilities are below the threshold for quasi-marginal stability of the corresponding $\alpha$TAE$(j,p)$ branch ($\alpha < \alpha_0^{(j,p)}$, as defined in section 3 of \cite{Bierwage10a}). The instabilities described in \cite{Hirose94} belong to this class [e.g., regions (IVa) and (V) in figure \ref{fig:scan-a_hirose}]. In the case of $\alpha$TAE above the threshold for quasi-marginal stability ($\alpha > \alpha_0^{(j,p)}$), continuum damping becomes negligible [$-\gamma_{\rm CD} \lesssim \O(10^{-4})$], so the instability thresholds in $\eta_{\rm i}$ and $\krhoiO$ are mainly due to Landau damping [e.g., regions (III) and (IVb) in figure \ref{fig:scan-a_dong}].

\paragraph{Non-resonant contribution.} Kinetic compression, ${\rm Re}(\delta W_{\rm ki})$, generally plays an important role in setting the eigenfrequency when the frequency is low enough so that the phase velocity of the wave becomes comparable to the typical ion sound speed, in which case part of the wave energy is spent on compressing and decompressing the ions. This effect was seen in the MHD ballooning unstable domain and in regions where the dominant $\alpha$TAE$(j,p)$ branch is below quasi-marginal stability ($\alpha < \alpha_0^{(j,p)}$); i.e., weakly trapped and subject to significant continuum damping when located outside a gap. Such weakly trapped $\alpha$TAEs couple particularly strongly to the ion sound branch when the perpendicular wavelength is large compared to the Larmor radius ($\krhoiO \lesssim 0.1$); i.e., near the instability threshold where diamagnetic effects are small. Examples were shown, where ${\rm Re}(\delta W_{\rm ki})$ rises above the ideal MHD potential energy $\delta W_{\rm f}$ as $\krhoiO$ decreases [see panels (b) and (c) in figures \ref{fig:scan-a_qforms_hirose} and \ref{fig:scan-a_qforms_dong}]. In one instance, an $\alpha$TAE$(2,0)$ is found to delve into the KTI gap, where continuum damping disappears [see figure \ref{fig:scaling_dong}(f)].

\paragraph{FLR effects.} As noted above, full FLR effects are important for AITG instabilities; in fact, in a self-consistent theory, FLR corrections and kinetic compression always appear together. In section 5.3 of \cite{Bierwage10a}, it was shown that the sharp accumulation points of the continuous spectrum known from lowest-order FLR models turn into smooth transitions when FLR effects are fully retained. One consequence is that the associated frequency shifts depend on the eigenmode structure. In \cite{Bierwage10a}, this phenomenon was discussed for the diamagnetic gap, but the same applies to the KTI gap (see Appendix).

\section{Conclusion}
\label{sec:conclude}

In this work, the linear destabilization of discrete shear Alfv\'{e}n eigenmodes in tokamak geometry has been studied numerically in the presence of kinetic thermal ion compression and thermal ion temperature gradient (ITG). The simulations where carried out using the 1-D linear $\delta f$ particle-in-cell code \textsc{awecs} \cite{Bierwage08}, which solves gyrokinetic equations based on a model by Chen \& Hasegawa \cite{Chen91} in a local flux tube and an $s$-$\alpha$ model equilibrium \cite{Connor78}. Of particular interest was the second magnetohydrodynamic (MHD) ballooning stable domain \cite{Coppi80}.

In the scenario considered, there are two fundamental types of ideal MHD shear Alfv\'{e}n eigenmodes induced by magnetic field non-uniformities due to curvature and a steep pressure gradient. The first is the well-known unstable ballooning mode (BM), for which ideal MHD potential energy is negative, $\delta W_{\rm f} < 0$. The second type encompasses the various branches of oscillatory ($\delta W_{\rm f} > 0$) $\alpha$-induced toroidal Alfv\'{e}n eigenmodes \cite{Hu04}. These so-called $\alpha$TAEs are standing waves trapped between pressure-gradient-induced potential barriers. In a companion paper \cite{Bierwage10a}, we described how $\alpha$TAEs are modified by finite-Larmor-radius (FLR) effects in the fluid limit ($\delta W_{\rm ki}=0$). In the present work, we dealt with the effects of kinetic compression, ($\delta W_{\rm ki} \neq 0$).

Our choice of parameters allowed us to revisit two cases which were examined in earlier numerical studies of Alfv\'{e}nic ITG (AITG) instabilities; one with lower magnetic shear, $s=0.4$ \cite{Hirose94}, and one with higher shear, $s=1.0$ \cite{Dong04}. The results reported by Hirose \textit{et al.} \cite{Hirose94} for the case with lower shear, $s=0.4$, are reproduced and we are now able to explain the observations in that and subsequent studies \cite{Hirose94, Dong04, Hirose95, Nordman95, Yamagiwa97, Joiner08} in terms of ITG-driven $\alpha$TAE ground states \cite{Hu04} and on the basis of the theoretical framework provided by the generalized fishbone-like dispersion relation (see \cite{Chen07} for a review). Apart from clarifying earlier results, this new interpretation has two important implications:
\begin{itemize}
\item  The eigenfrequencies of the AITG instabilities in the second MHD ballooning stable domain are determined by the frequency of the ideal MHD $\alpha$TAE, which is modified by FLR effects \cite{Bierwage10a} and coupling to ion sound waves (section \ref{sec:qforms}).

\item  The finite instability thresholds in temperature gradient ($\eta_{\rm i}$) and wavenumber ($\krhoiO$) are due to resonant absorption by both thermal ions (Landau damping) and shear Alfv\'{e}n waves (continuum damping). Both absorption rates can be comparable, except for quasi-marginally stable $\alpha$TAEs, for which continuum damping is negligible.
\end{itemize}

Qualitatively similar results are obtained in the case with higher shear, $s=1.0$: here the second MHD stable domain is also destabilized by ITG-driven $\alpha$TAEs in the entire parameter range scanned. Thus, our results disagree with those presented by Dong \textit{et al.} \cite{Dong04} who observed a destabilization only near the second MHD stability boundary.

Quasi-marginally stable $\alpha$TAEs ($\alpha > \alpha_0$, as defined in \cite{Bierwage10a}) seem to be difficult to excite by interaction with thermal ions: the resulting AITG instabilities have relatively low growth rates and are often sub-dominant. This is presumably due to their high frequencies $\omega_{\rm r} \gg \omega_{*p{\rm i}}$ due to strong field line bending. The excitation of these modes in the presence of an additional population of energetic ions has been demonstrated with an ideal-MHD-gyrokinetic hybrid model \cite{Hu05}. Work is underway to study $\alpha$TAE excitation with both gyrokinetic thermal and energetic ions. The destabilization of shear Alfv\'{e}n waves in the presence of an electron temperature gradient (ETG) \cite{Dong04, Joiner07} may also be revisited and explored in the light of $\alpha$TAEs.

The results of this and related earlier studies indicate that the resonant excitation of $\alpha$TAEs may potentially destabilize a large part of the second MHD ballooning stable domain, including the domain of negative magnetic shear. However, since the properties of $\alpha$TAEs crucially depend on the structure of the SAW Schr\"{o}dinger potential, the present results can only serve as a proof-of-principle to motivate further work in this direction. Practically relevant predictions regarding the robustness of the second MHD stable domain with respect to kinetic instabilities require further study employing more realistic high-$\beta$ equilibrium models (such as in \cite{Miller98}) and, eventually, global nonlinear simulations. In this context, let us note that the radial mode structures of local $\alpha$TAE solutions have a half-width extending to neighboring rational surfaces ($nq-m = \pm 1$), which indicates that these modes experience strong toroidal coupling. This suggests that, provided the modes can be excited to significant nonlinear amplitudes, they may contribute to the radial transports in high-$\beta$ toroidal devices.

The present analysis of properties and structures of Alfv\'{e}nic fluctuations in the presence of steep pressure gradients applies for both positive or negative magnetic shear and can serve as an interpretative framework for experimental observations in (future) high-performance fusion plasmas of reactor relevance.

\ack

One of the authors (A.B.) would like to thank Shuanghui Hu (Guizhou University) for helpful advice during the early stages of the project. He also thanks Masaru Furukawa (Tokyo University) and Zhihong Lin (UCI) for stimulating discussions. This research is supported by U.S.~DoE Grant DE-AC02-CH0-3073, NSF Grant ATM-0335279, and SciDAC GSEP.

\appendix
\section*{Appendix: FLR effects on the kinetic thermal ion (KTI) gap}
\setcounter{section}{1}
\label{sec:appendix_soundgap}

In the limit $|\krhoiO s \theta| \gg 1$, equation (\ref{eq:model_saw_red}) asymptotically approaches
\begin{equation}
0 = \delta\Psi_{\rm s}'' + \frac{X(\omega)}{(\krhoiO s \theta)^2} \delta\Psi_{\rm s} + Y(\omega) \frac{\sin\theta}{\krhoiO s \theta} \delta\Psi_{\rm s}, \qquad (|\krhoiO s \theta| \gg 1);
\label{eq:saw_eq_flr_bc}
\end{equation}

\noindent where
\begin{equation}
\fl X(\omega) = (\omega + \tauei^T\omega_{*{\rm i}})(\omega - \omega_{*{\rm i}}) / (1 + \tauei^T), \quad
Y(\omega) = (\omega - \omega_{*p{\rm i}}) 2 q \vti
\label{eq:saw_eq_flr_bc_coeff}
\end{equation}

\noindent Equation (\ref{eq:saw_eq_flr_bc}) describes FLR continuum waves and is the same as that obtained for the FLR MHD model \cite{Bierwage10a} because $\delta G_{\rm i}$ only contributes a negligible term of the form $\delta\Psi_{\rm s}\sin(\theta)/\theta^2$. This may be readily verified by noting that, in the limit $|\krhoiO s\theta| \gg 1$, the gyrokinetic Vlasov equation (\ref{eq:model_vlasov_gke2}) reduces to $\delta G_{\rm i} \approx (e_{\rm i}/\mi) Q_{\rm i}\feqi J_0 \delta\psi$ and enters primarily via the term $\left<\omegad J_0 \delta G\right>_{\rm i}$ in the vorticity equation (\ref{eq:model_maxw_vort1}).

Equation (\ref{eq:saw_eq_flr_bc}) was solved and discussed in detail in the Appendix of \cite{Bierwage10a}. The discussion of the diamagnetic frequency shift in section 5.3 of \cite{Bierwage10a} illustrates how the accumulation points of the shear Alfv\'{e}n continuum becomes blurred when FLR effects are important. The same argument can now be applied in a discussion on how FLR effects affect the accumulation point of the kinetic thermal ion (KTI) gap.

When FLR effects are important, the amount of energy an eigenmode transfers to an outgoing continuum wave depends on its eigenfrequency, $\omega_{\rm r}$, as well as on its mode structure; i.e., localization along $\theta$. The continuum damping disappears only gradually when the mode frequency drops below the nominal value of the upper accumulation point of the KTI gap obtained in the low-$b_{\rm i}$ limit. In fact, since terms accounting for kinetic compression become negligible in the large-$|\krhoiO s\theta|$ limit [equation (\ref{eq:saw_eq_flr_bc})], the lower bound of the KTI gap width is \textit{zero} in the present model. Nevertheless, kinetic compression ensures that any regular eigenmode (finite $k_\parallel$) experiences only reduced continuum damping.

\setlength{\bibsep}{0.6pt}
\bibliographystyle{unsrt}

\end{document}